\documentclass
[preprint,prd,secnumarabic,showkeys,showpacs,nofootinbib,hyperref]{revtex4}%
\usepackage{amsfonts}
\usepackage{amsmath}
\usepackage{amssymb}
\usepackage{graphicx}%
\providecommand{\U}[1]{\protect\rule{.1in}{.1in}}

\begin{document}
\title{Path Integral Quantization of Noncommutative Complex Scalar Field}
\author{Farid Khelili}
\affiliation{20 Aout 55 Skikda University, Skikda, Algeria}
\keywords{Noncommutative space, Noncommutative scalar field, Path integral, Propagators,
Renormalization, Dimensional regularization.}
\pacs{}

\begin{abstract}
Using noncommutative deformed canonical commutation relations, a model
describing a noncommutative complex scalar field theory is considered. Using
the path integral formalism, the noncommutative free and exact propagators are
calculated to one-loop order and to the second order in the parameter of
noncommutativity. Dimensional regularization was used to remove ultraviolet
divergences that arise from loop graphs. It has been shown that these
divergences may also be absorbed into a redefinition of the parameters of the theory.

\end{abstract}
\maketitle

\section{Introduction}

Noncommutative geometry is presently one of the most important and extremely
active area of research in theoretical physics, there is now a common belief
that the usual picture of space-time as a smooth pseudo-Riemannian manifold
should breakdown at very short distances of the order of the Planck length,
due to the quantum gravity effects. The concept of noncommutative space-time
was suggested very early on by the founding fathers of quantum mechanics and
quantum field theory. This was motivated by the need to remove the divergences
which had plagued quantum electrodynamics. However, this suggestion was
ignored \cite{NCS1} . In recent years, the idea of noncommutative space-time
has attracted considerable interest, and has penetrated into various fields in
physics and mathematical physics, it leads to the investigation of some new
and more fundamental physical and mathematical notions, the motivation for
this kind of investigation is that the effects of noncommutativity of space
may appear at very short distances of the order of the Planck length, or at
very high energies, this may shed a light on the real microscopic geometry and
structure of our universe \cite{NCS1}-\cite{NCS14}. One of the new features of
noncommutative field theories is the UV/IR mixing phenomenon, in which the
physics at high energies affects the physics at low energies, which does not
occur in quantum field theories in which the coordinates commute \cite{NCS1}
\cite{NCS12} \cite{NCS13}. Thus in the noncommutative space-time approach, the
dynamical variables become operators, and, therefore, the formalism of the
quantum field theory constructions must be modified. The discovery of
noncommutative geometry has allowed the exploration of new directions in
theoretical physics, in particular, several aspects of noncommutative quantum
mechanics and two-dimensional noncommutative harmonic oscillators are an
extremely active area of research and have been discussed extensively from
different points of view \cite{OSC1}-\cite{Saha3}.

Our paper is organized as follows: In Section 2, we consider a noncommutative
action for a complex scalar field with self interaction, in section 3, we
briefly recall basic results of path integral derivation of propagator and
renormalization of $\varphi^{4}-$ theory, in section 4, we consider the path
integral derivation of the noncommutative free and exact propagators, in
section 5 we consider the dimensional regularization of the noncommutative
exact propagator, Finally in section 6, we draw our conclusions.

\section{Noncommutative Action}

In \cite{Khelili} a model describing a noncommutative complex scalar field
theory, based on noncommutative deformed canonical commutation relations, was
considered, the model has been quantized via the Peierls bracket \cite{Witt}.
Here we will use the path integral formalism to quantize the noncommutative
complex scalar field theory.

Consider a complex scalar field $\Phi\left(  x\right)  $ with Lagrangian
density given by \cite{WEINB}-\cite{BRN}%

\begin{equation}
\mathfrak{L}=-\left(  \partial_{\mu}\Phi\right)  ^{\ast}\left(  \partial^{\mu
}\Phi\right)  -m^{2}\Phi^{\ast}\Phi-g\left(  \Phi^{\ast}\Phi\right)  ^{2}
\label{eqn1}%
\end{equation}

where $m$ is the mass of the charged particles, and $g$ is a positive
parameter. The metric signature will be assumed to be $-++$..., in what
follows, we take $\hbar=c=1$.

The complex scalar field can be quantized using the canonical quantization
rules, for this we express it in terms of its real and imaginary parts as
$\Phi=\frac{1}{\sqrt{2}}\left(  \varphi_{1}+i\varphi_{2}\right)  $, where
$\varphi_{1},\varphi_{2}$ are real scalar fields; in terms of these real
scalar fields the Lagrangian density reads%

\begin{equation}
\mathfrak{L}=-\frac{1}{2}\left(  \partial_{\mu}\varphi_{a}\right)  ^{2}%
-\frac{1}{2}m^{2}\left(  \varphi_{a}\right)  ^{2}-\frac{1}{4}g\left(
\varphi_{a}\varphi_{a}\right)  ^{2}=-\frac{1}{2}\left(  \partial_{\mu}%
\varphi_{a}\right)  ^{2}-\frac{1}{2}\mu^{2}\left[  \varphi\right]  \left(
\varphi_{a}\right)  ^{2} \label{eqn2}%
\end{equation}

where $\mu^{2}\left[  \varphi\right]  =m^{2}+\frac{1}{2}g\left(  \varphi
_{a}\right)  ^{2}.$

Let $\pi_{a}$ be the canonical conjugate to $\varphi_{a}$%
\begin{equation}
\pi_{a}=\frac{\partial\mathfrak{L}}{\partial\overset{.}{\varphi}_{a}}%
=\overset{\cdot}{\varphi}_{a} \label{eqn3}%
\end{equation}

The Hamiltonian density reads then
\begin{equation}
\mathcal{H}=\pi_{a}\overset{\cdot}{\varphi}_{a}-\mathfrak{L}\mathcal{=}%
\frac{1}{2}\left(  \pi_{a}\right)  ^{2}+\frac{1}{2}\left(  \overrightarrow
{\nabla}\varphi_{a}\right)  ^{2}+\frac{1}{2}\mu^{2}\left[  \varphi\right]
\left(  \varphi_{a}\right)  ^{2} \label{eqn4}%
\end{equation}

In the canonical quantization the canonical variables $\varphi_{a}$ and the
canonical conjugates $\pi_{a}$ are assumed to be operators satisfying the
canonical commutation relations%

\begin{align}
\left[  \varphi_{a}\left(  t,\overrightarrow{x}\right)  ,\pi_{b}\left(
t,\overrightarrow{y}\right)  \right]   &  =i\delta_{ab}\delta^{3}\left(
\overrightarrow{x}-\overrightarrow{y}\right) \label{eqn5}\\
\left[  \varphi_{a}\left(  t,\overrightarrow{x}\right)  ,\varphi_{b}\left(
t,\overrightarrow{y}\right)  \right]   &  =0\nonumber\\
\left[  \pi_{a}\left(  t,\overrightarrow{x}\right)  ,\pi_{b}\left(
t,\overrightarrow{y}\right)  \right]   &  =0\nonumber
\end{align}

It is well known, since the birth of quantum field theory in the papers of
Born, Dirac, Fermi, Heisenberg, Jordan, and Pauli, that the free field behaves
like an infinite number of coupled harmonic oscillators \cite{WEINB}, using
this analogy between free fields and an infinite number of coupled harmonic
oscillators, one can impose non commutativity on the configuration space of
dynamical fields $\varphi_{a}$, to do this we recall that the two-dimensional
harmonic oscillator noncommutative configuration space can be realized as a
space where the coordinates $\widehat{x}_{a}$, and the corresponding
noncommutative momentum $\widehat{p}_{a},$ are operators satisfying the
commutation relations%

\begin{equation}
\left[  \widehat{x}_{a},\widehat{x}_{b}\right]  =i\theta^{2}\varepsilon
_{ab}\ \ \ \ \ \left[  \widehat{p}_{a},\widehat{p}_{b}\right]
=0\ \ \ \ \ \left[  \widehat{x}_{a},\widehat{p}_{b}\right]  =i\delta_{ab}
\label{eqn5a}%
\end{equation}

where $\theta$ is a parameter with dimension of length, and $\varepsilon_{ab}$
is an antisymmetric constant matrix.\bigskip

It is well known that this noncommutative algebra can be mapped to the
commutative Heisenberg-Weyl algebra \cite{Saha1}-\cite{Saha3}
\begin{equation}
\left[  x_{a},x_{b}\right]  =0\ \ \ \ \ \left[  p_{a},p_{b}\right]
=0\ \ \ \ \ \left[  x_{a},p_{b}\right]  =i\delta_{ab} \label{eqn5b}%
\end{equation}

through the relations%

\begin{equation}
\widehat{x}_{a}=x_{a}-\frac{1}{2}\theta^{2}\varepsilon_{ab}p_{b}%
\ \ \ \ \ \widehat{p}_{a}=p_{a} \label{eqn5c}%
\end{equation}

To impose non commutativity on the configuration space of dynamical fields
$\varphi_{a}$, we assume that the noncommutative canonical variables
$\widehat{\varphi}_{a}$ and the noncommutative canonical conjugates
$\widehat{\pi}_{a}$ satisfy the noncommutative commutation relations%

\begin{align}
\left[  \widehat{\varphi}_{a}\left(  t,\overrightarrow{x}\right)
,\widehat{\pi}_{b}\left(  t,\overrightarrow{y}\right)  \right]   &
=i\delta^{3}\left(  \overrightarrow{x}-\overrightarrow{y}\right)  \delta
_{ab}\label{eqn6}\\
\left[  \widehat{\varphi}_{a}\left(  t,\overrightarrow{x}\right)
,\widehat{\varphi}_{b}\left(  t,\overrightarrow{y}\right)  \right]   &
=i\theta\varepsilon_{ab}\delta^{3}\left(  \overrightarrow{x}-\overrightarrow
{y}\right) \nonumber\\
\left[  \widehat{\pi}_{a}\left(  t,\overrightarrow{x}\right)  ,\widehat{\pi
}_{b}\left(  t,\overrightarrow{y}\right)  \right]   &  =0\nonumber
\end{align}

where $\theta$ is the parameter of noncommutativity, which is assumed to be a
constant, and $\varepsilon_{ab}$ is a $2\times2$ real antisymmetric matrix%

\begin{equation}
\varepsilon_{12}=-\varepsilon_{21}=1 \label{eqn7}%
\end{equation}

The noncommutative Hamiltonian density is assumed to have the form%

\begin{equation}
\widehat{\mathcal{H}}\mathcal{=}\frac{1}{2}\left(  \widehat{\pi}_{a}\right)
^{2}+\frac{1}{2}\left(  \overrightarrow{\nabla}\widehat{\varphi}_{a}\right)
^{2}+\frac{1}{2}\mu^{2}\left[  \widehat{\varphi}\right]  \left(
\widehat{\varphi}_{a}\right)  ^{2} \label{eqn8}%
\end{equation}

By generalizing the noncommutative harmonic oscillator construction\ an
extension of quantum field theory based on the concept of noncommutative
fields satisfying the noncommutative commutation relations $\left(
\ref{eqn6}\right)  $ has been proposed in \cite{OSC13}-\cite{OSC15}, where the
properties and phenomenological implications of the noncommutative field has
been studied and applied to different problem including scalar, gauge and
fermionic fields \cite{OSC13}-\cite{OSC17}. Our approach is different, it is
based on the relation between the noncommutative variables $\widehat{\varphi
}_{a}$ and $\widehat{\pi}_{a}$ and the canonical variables $\varphi_{a}$ and
$\pi_{a}$.

It is easy to see that the noncommutative commutation relations $\left(
\ref{eqn6}\right)  $ can be mapped to the canonical commutation relations
$\left(  \ref{eqn5}\right)  $ if the noncommutative variables $\widehat
{\varphi}_{a}$ and $\widehat{\pi}_{a}$ are related to the canonical variables
$\varphi_{a} $ and $\pi_{a}$ by the relations%

\begin{align}
\widehat{\varphi}_{a}  &  =\varphi_{a}-\frac{1}{2}\theta\varepsilon_{ab}%
\pi_{b}\label{eqn9}\\
\widehat{\pi}_{a}  &  =\pi_{a}\nonumber
\end{align}

Using these transformations, the noncommutative Hamiltonian density eq$\left(
\ref{eqn8}\right)  $ can be rewritten, up to a total derivative term and up to
second order in the parameter $\theta$, as%

\begin{equation}
\widehat{\mathcal{H}}\mathcal{=}\frac{1}{2}\pi^{\sim}\mathbb{M}\pi-\frac{1}%
{8}\theta^{2}\pi^{\sim}\mathbb{D}\pi+\theta\pi^{\sim}\mathbb{N\varphi+}%
\frac{1}{2}\varphi^{\sim}\mathbb{B}\varphi+O\left(  \theta^{3}\right)
\label{eqn10}%
\end{equation}

where%

\begin{gather}
\mathbb{M}\mathbb{=I+}\frac{1}{4}\theta^{2}\left(  m^{2}\mathbb{I}%
-g\varepsilon\widehat{\sigma}\varepsilon\right)  =\mathbb{M}^{\sim
}\label{eqn11}\\
\widehat{\sigma}_{ab}=\frac{\delta^{2}}{\delta\varphi_{a}\delta\varphi_{b}%
}\left[  \frac{1}{4}\left(  \varphi^{\sim}\varphi\right)  ^{2}\right]
=\varphi^{\sim}\varphi\delta_{ab}+2\varphi_{a}\varphi_{b}\nonumber\\
\widehat{\sigma}=\left(  \varphi^{\sim}\varphi\right)  \mathbb{I}%
+2\mathit{M}\left[  \varphi\right]  \text{ \ , \ }\mathit{M}_{ab}\left[
\varphi\right]  \text{ }=\varphi_{a}\varphi_{b}\nonumber\\
\mathbb{D}\mathbb{=}\overrightarrow{\mathbb{\nabla}}^{2}\mathbb{I=D}^{\sim
}\nonumber\\
\mathbb{N}\mathbb{=}\frac{1}{2}\left(  m^{2}-\overrightarrow{\mathbb{\nabla}%
}^{2}+g\left(  \varphi_{a}\right)  ^{2}\right)  \varepsilon=-\mathbb{N}^{\sim
}\nonumber\\
\mathbb{B}\mathbb{=}\left(  m^{2}-\overrightarrow{\mathbb{\nabla}}^{2}%
+\frac{1}{2}g\left(  \varphi_{a}\right)  ^{2}\right)  \mathbb{I}%
=\mathbb{B}^{\sim}\nonumber
\end{gather}

with $\mathbb{I}$\ denotes the $2\times2$ unit matrix, and $\mathbb{A}^{\sim}$
denotes the transpose of the operator $\mathbb{A}.$

From now on we keep only the modifications due to the noncommutativity up to
second order in the parameter $\theta.$

The relation between $\pi_{a}$ and $\overset{\cdot}{\varphi}_{a}$ is given by%

\begin{equation}
\overset{\cdot}{\varphi}_{a}\left(  x\right)  =\frac{\delta\widehat{H}}%
{\delta\pi_{a}\left(  x\right)  } \label{eqn12}%
\end{equation}

where $\widehat{H}=\int d^{3}x\widehat{\mathcal{H}}.$ Using the expression of
$\widehat{\mathcal{H}}$ and the symmetry properties of the operators
$\mathbb{M=M}^{\sim}$ and $\mathbb{D=D}^{\sim}$, one gets%

\begin{equation}
\overset{\cdot}{\varphi}_{a}\left(  x\right)  =\mathbb{M}_{ab}\pi_{b}\left(
x\right)  -\frac{1}{4}\theta^{2}\mathbb{D}_{ab}\pi_{b}\left(  x\right)
+\theta\mathbb{N}_{ab}\mathbb{\varphi}_{b}\left(  x\right)  \label{eqn13}%
\end{equation}

From this relation we get the following iterative expression of $\pi_{a}$%

\begin{equation}
\pi_{a}=\mathbb{K}_{ab}\left(  \overset{\cdot}{\varphi}_{b}-\theta
\mathbb{N}_{bc}\mathbb{\varphi}_{c}\right)  +\frac{1}{4}\theta^{2}%
\mathbb{K}_{ab}\mathbb{D}_{bc}\pi_{c} \label{eqn14}%
\end{equation}

where $\mathbb{K}$ is the inverse of the matrix $\mathbb{M}$%

\begin{equation}
\mathbb{K=M}^{-1}=\mathbb{I-}\frac{1}{4}\theta^{2}\left(  m^{2}\mathbb{I}%
-g\varepsilon\widehat{\sigma}\varepsilon\right)  \label{eqn15}%
\end{equation}

Using the expression of the matrix $\mathbb{K}$ one gets, by iteration, the
following expression of $\pi_{a}$%
\begin{equation}
\pi_{a}=\left(  \mathbb{I+}\frac{1}{4}\theta^{2}\overline{\mathbb{D}}\right)
_{ab}\overset{\cdot}{\varphi}_{b}-\theta\mathbb{N}_{ab}\mathbb{\varphi}_{b}
\label{eqn16}%
\end{equation}

where $\overline{\mathbb{D}}=\mathbb{D-}\left(  m^{2}\mathbb{I}-\varepsilon
\widehat{\sigma}\varepsilon\right)  =\overrightarrow{\mathbb{\nabla}}%
^{2}\mathbb{I}-\left(  m^{2}\mathbb{I}-g\varepsilon\widehat{\sigma}%
\varepsilon\right)  =\overline{\mathbb{D}}^{\sim}.$

We note that the noncommutative Hamiltonian density can be derived from the
following noncommutative Lagrangian density%

\begin{equation}
\widehat{\mathfrak{L}}=\frac{1}{2}\overset{\cdot}{\varphi^{\sim}}\left(
\mathbb{I+}\frac{1}{4}\theta^{2}\overline{\mathbb{D}}\right)  \overset{\cdot
}{\varphi}+\theta\mathbb{\varphi}^{\sim}\mathbb{N}\overset{\cdot}{\varphi
}-\frac{1}{2}\mathbb{\varphi}^{\sim}\left(  \mathbb{B+}\theta^{2}%
\mathbb{N}^{2}\right)  \varphi\label{eqn17}%
\end{equation}

via the usual Legendre transformation $\widehat{\mathfrak{L}}=\pi_{a}%
\overset{\cdot}{\varphi}_{a}-$ $\widehat{\mathcal{H}}$. To get this expression
we have used the symmetry properties of the operators $\overline{\mathbb{D}}$,
$\mathbb{N}$ and $\mathbb{B}.$

\section{Propagator and Renormalization}

In this section we briefly recall basic results of path integral derivation of
propagators and renormalization of $\varphi^{4}-$ theory \cite{BRN} \cite{MAG}
\cite{WEINB} \cite{QFT1}-\cite{QFT4}.

Let us consider the theory of a real self-interacting scalar field specified
by an action of the form%

\begin{equation}
\mathcal{S}\left[  \varphi\right]  =\int d^{4}x\left[  -\frac{1}{2}\left[
\partial_{\mu}\varphi\left(  x\right)  \right]  ^{2}-\frac{1}{2}m^{2}\left[
\varphi\left(  x\right)  \right]  ^{2}-\frac{1}{4!}g\left[  \varphi\left(
x\right)  \right]  ^{4}\right]  \label{eqn a9}%
\end{equation}

The action can be written as%

\begin{equation}
\mathcal{S}\left[  \varphi\right]  =\mathcal{S}^{\left(  0\right)  }\left[
\varphi\right]  +\mathcal{S}^{\left(  1\right)  }\left[  \varphi\right]
\end{equation}

where $\mathcal{S}^{\left(  0\right)  }\left[  \varphi\right]  $ is the free
action
\begin{equation}
\mathcal{S}^{\left(  0\right)  }\left[  \varphi\right]  =-\frac{1}{2}\int
d^{4}xd^{4}y\mathbb{\varphi}\left(  x\right)  \mathcal{D}\left(  x,y\right)
\mathbb{\varphi}\left(  y\right)
\end{equation}

and $\mathcal{S}^{\left(  1\right)  }\left[  \varphi\right]  $ is the
interaction term
\begin{equation}
\mathcal{S}^{\left(  1\right)  }\left[  \varphi\right]  =-\frac{1}{4!}g\int
d^{4}x\left[  \varphi\left(  x\right)  \right]  ^{4}%
\end{equation}

with%

\begin{equation}
\mathcal{D}\left(  x,y\right)  =\left(  m^{2}-\square\right)  \delta
^{4}\left(  x,y\right)  =\int\frac{d^{4}q}{\left(  2\pi\right)  ^{4}}\left(
m^{2}+q^{2}\right)  e^{iq\left(  x-y\right)  }=\mathcal{D}\left(  y,x\right)
\end{equation}

The free propagator $\Delta\left(  x,y\right)  $ is defined as the vacuum
expectation value \cite{WEINB} \cite{QFT1}-\cite{QFT4}
\begin{equation}
-i\Delta\left(  x,y\right)  =\left\langle 0\right\vert T\left\{  \Phi\left(
x\right)  \Phi\left(  y\right)  \right\}  \left\vert 0\right\rangle
\label{eqn a1}%
\end{equation}

where $\left\langle 0\right\vert T\left\{  \mathcal{A}\left[  \Phi\right]
\right\}  \left\vert 0\right\rangle $\ denotes the vacuum expectation value of
the chronological product $T\left\{  \mathcal{A}\left[  \Phi\right]  \right\}
$, $\Phi\left(  x\right)  $ denotes the free quantum field operator
corresponding to $\mathbb{\varphi}\left(  x\right)  $, and $T$ is the
time-ordering operator.

The vacuum expectation value eq$\left(  \ref{eqn a1}\right)  $ can be
expressed as a Feynman functional integral \cite{WEINB} \cite{QFT1}
\cite{QFT2}
\begin{equation}
-i\Delta\left(  x,y\right)  =\left\langle 0\right\vert T\left\{  \Phi\left(
x\right)  \Phi\left(  y\right)  \right\}  \left\vert 0\right\rangle
=\frac{\int%
{\displaystyle\prod\limits_{x}}
d\mathbb{\varphi}\left(  x\right)  \left[  \mathbb{\varphi}\left(  x\right)
\mathbb{\varphi}\left(  y\right)  \right]  e^{i\mathcal{S}^{\left(  0\right)
}\left[  \varphi\right]  }}{\int%
{\displaystyle\prod\limits_{x}}
d\mathbb{\varphi}\left(  x\right)  e^{i\mathcal{S}^{\left(  0\right)  }\left[
\varphi\right]  }} \label{eqn a4}%
\end{equation}

The functional integrals in eq$\left(  \ref{eqn a4}\right)  $ can now be
performed with the help of the identity \cite{WEINB}%

\begin{equation}
\int%
{\displaystyle\prod\limits_{x}}
d\mathbb{\varphi}\left(  x\right)  \left[  \mathbb{\varphi}\left(  x_{l_{1}%
}\right)  \mathbb{\varphi}\left(  x_{l_{2}}\right)  ...\mathbb{\varphi}\left(
x_{2N}\right)  \right]  e^{i\mathcal{S}^{\left(  0\right)  }\left[
\varphi\right]  }=\left[  \det\left(  \frac{i\mathcal{D}}{2\pi}\right)
\right]  ^{-\frac{1}{2}}\sum_{\substack{pairings\\of\ \left(  l_{1}%
,l_{2},...,l_{2N}\right)  }}%
{\displaystyle\prod\limits_{pairs\left(  l_{j},l_{k}\right)  }}
\left[  -i\mathcal{D}^{-1}\left(  x_{l_{j}},x_{l_{k}}\right)  \right]
\label{eqn a7}%
\end{equation}

where the sum is over all ways of pairing the indices $l_{1},l_{2},...,l_{2N}%
$, with two pairings being considered the same if they differ only by the
order of the pairs, or by the order of indices within a pair, and
$\mathcal{D}^{-1}\left(  x,y\right)  $ is the inverse of the matrix
$\mathcal{D}\left(  x,y\right)  $.

Direct calculations give the following expression for the propagator%
\begin{equation}
-i\Delta\left(  x,y\right)  =-i\mathcal{D}^{-1}\left(  x,y\right)
\label{eqn 37-f}%
\end{equation}

to get the inverse of the matrix $\mathcal{D}\left(  x,y\right)  $ we write
$\mathcal{D}\left(  x,y\right)  $ as a Fourier integral%

\begin{equation}
\mathcal{D}\left(  x,y\right)  =\int\frac{d^{4}q}{\left(  2\pi\right)  ^{4}%
}\mathcal{D}\left(  q\right)  e^{iq\left(  x-y\right)  }%
\end{equation}

where%
\begin{equation}
\mathcal{D}\left(  q\right)  =m^{2}+q^{2}-i\epsilon
\end{equation}

Eq $\left(  \ref{eqn 37-f}\right)  $\ can be rewritten as an integral equation%

\begin{equation}
\int d^{4}y\mathcal{D}\left(  x,y\right)  \Delta\left(  y,z\right)
=\delta\left(  x,z\right)  \label{eqn 37-g}%
\end{equation}

The solution of eq $\left(  \ref{eqn 37-g}\right)  $ is then%
\begin{equation}
\Delta\left(  x,y\right)  =\int\frac{d^{4}q}{\left(  2\pi\right)  ^{4}}%
\Delta\left(  q\right)  e^{iq\left(  x-y\right)  }%
\end{equation}

where $\Delta\left(  q\right)  =\mathcal{D}^{-1}\left(  q\right)  =\frac
{1}{m^{2}+q^{2}-i\epsilon}$ is the free-field propagator, and the $i\epsilon$
terms have the effect of making the inverse well-defined for all real values
of $q$.

The exact propagator $\Delta^{\prime}\left(  x,y\right)  $ is given by
\cite{WEINB} \cite{QFT1} \cite{QFT2}%

\begin{equation}
-i\Delta^{\prime}\left(  x,y\right)  =\left\langle 0\right\vert T\left\{
\Phi\left(  x\right)  \Phi\left(  y\right)  \right\}  \left\vert
0\right\rangle =\frac{\int%
{\displaystyle\prod\limits_{x}}
d\mathbb{\varphi}\left(  x\right)  \left[  \mathbb{\varphi}\left(  x\right)
\mathbb{\varphi}\left(  y\right)  \right]  e^{i\mathcal{S}\left[
\varphi\right]  }}{\int%
{\displaystyle\prod\limits_{x}}
d\mathbb{\varphi}\left(  x\right)  e^{i\mathcal{S}\left[  \varphi\right]  }}
\label{eqn a2}%
\end{equation}

In order to evaluate the functional integral eq$\left(  \ref{eqn a2}\right)
$, we make a Taylor expansion in powers of g\ %

\begin{equation}
e^{i\mathcal{S}\left[  \varphi\right]  }=e^{i\mathcal{S}^{\left(  0\right)
}\left[  \varphi\right]  +i\mathcal{S}^{\left(  1\right)  }\left[
\varphi\right]  }=e^{i\mathcal{S}^{\left(  0\right)  }\left[  \varphi\right]
}\left(  1+i\mathcal{S}^{\left(  1\right)  }\left[  \varphi\right]
+...\right)
\end{equation}

Using this Taylor expansion in eq$\left(  \ref{eqn a2}\right)  $ we obtain the
following expressions%

\begin{align}
\int%
{\displaystyle\prod\limits_{x}}
d\mathbb{\varphi}\left(  x\right)  \left[  \mathbb{\varphi}\left(  x\right)
\mathbb{\varphi}\left(  y\right)  \right]  e^{i\mathcal{S}\left[
\varphi\right]  }  &  =\int%
{\displaystyle\prod\limits_{x}}
d\mathbb{\varphi}\left(  x\right)  \left[  \mathbb{\varphi}\left(  x\right)
\mathbb{\varphi}\left(  y\right)  \right]  e^{i\mathcal{S}^{\left(  0\right)
}\left[  \varphi\right]  }\label{eqn a5}\\
&  -\frac{i}{4!}g\int d^{4}z\int%
{\displaystyle\prod\limits_{x}}
d\mathbb{\varphi}\left(  x\right)  \left[  \mathbb{\varphi}\left(  x\right)
\mathbb{\varphi}\left(  y\right)  \left[  \varphi\left(  z\right)  \right]
^{4}\right]  e^{i\mathcal{S}^{\left(  0\right)  }\left[  \varphi\right]
}\nonumber
\end{align}

and%

\begin{equation}
\int%
{\displaystyle\prod\limits_{x}}
d\mathbb{\varphi}\left(  x\right)  e^{i\mathcal{S}\left[  \varphi\right]
}=\int%
{\displaystyle\prod\limits_{x}}
d\mathbb{\varphi}\left(  x\right)  e^{i\mathcal{S}^{\left(  0\right)  }\left[
\varphi\right]  }-\frac{i}{4!}g\int d^{4}z\int%
{\displaystyle\prod\limits_{x}}
d\mathbb{\varphi}\left(  x\right)  \left[  \varphi\left(  z\right)  \right]
^{4}e^{i\mathcal{S}^{\left(  0\right)  }\left[  \varphi\right]  }
\label{eqn a6}%
\end{equation}

The functional integrals in eq$\left(  \ref{eqn a5}\right)  $ and eq$\left(
\ref{eqn a6}\right)  $\ can now be performed with the help of the identity
eq$\left(  \ref{eqn a7}\right)  $, the exact propagator $\Delta^{\prime
}\left(  x,y\right)  $ can be written as

\begin{equation}
\Delta^{\prime}\left(  x,y\right)  =\int\frac{d^{4}q}{\left(  2\pi\right)
^{4}}\Delta^{\prime}\left(  q\right)  e^{iq\left(  x-y\right)  }%
\end{equation}

where%

\begin{equation}
\Delta^{\prime}\left(  q\right)  =\Delta\left(  q\right)  +\Delta\left(
q\right)  \left[  \Pi_{Loop}^{\ast}\left(  q\right)  \right]  \Delta\left(
q\right)
\end{equation}

and $\Pi_{Loop}^{\ast}\left(  q\right)  $ is given by the divergent integral%

\begin{equation}
\Pi_{Loop}^{\ast}\left(  q\right)  =\frac{i}{2}g\int\frac{d^{4}p}{\left(
2\pi\right)  ^{4}}\frac{1}{m^{2}+p^{2}-i\epsilon} \label{eqn a8}%
\end{equation}

such ultraviolet divergences are typical of loop graphs. To deal with these
divergent integrals that appear in quantum field theory, one can use some sort
of regularization technique that makes these integrals finite. Dimensional
regularization is the most convenient method for regulating divergent
integrals, the idea is to treat the loop integrals as integrals over
D-dimensional momenta, and then take the limit $D\rightarrow4$, it turns out
that the singularity of 1-loop graphs are simple poles in $D=4$ \cite{BRN}
\cite{MAG} \cite{WEINB} \cite{Collins} \cite{QFT1}-\cite{QFT4}.

First we generalize the 4-dimensional action eq$\left(  \ref{eqn a7}\right)  $
to D-dimensions \cite{WEINB} \cite{QFT2}%

\begin{equation}
\mathcal{S}\left[  \varphi_{B}\right]  =\mathcal{S}_{B}^{\left(  0\right)
}\left[  \varphi_{B}\right]  +\mathcal{S}_{B}^{\left(  1\right)  }\left[
\varphi_{B}\right]
\end{equation}

with%

\begin{align}
\mathcal{S}_{B}^{\left(  0\right)  }\left[  \varphi_{B}\right]   &  =-\frac
{1}{2}\int d^{D}xd^{D}y\mathbb{\varphi}_{B}\left(  x\right)  \mathcal{D}%
^{\left(  B\right)  }\left(  x,y\right)  \mathbb{\varphi}_{B}\left(  y\right)
\\
\mathcal{D}^{\left(  B\right)  }\left(  x,y\right)   &  =\left(  m_{B}%
^{2}-\square\right)  \delta^{4}\left(  x,y\right)  =\int\frac{d^{D}q}{\left(
2\pi\right)  ^{D}}\left(  m_{B}^{2}+q^{2}\right)  e^{iq\left(  x-y\right)  }%
\end{align}

and%
\begin{equation}
\mathcal{S}_{B}^{\left(  1\right)  }\left[  \varphi_{B}\right]  =-\frac{1}%
{4!}\mu^{4-D}g_{B}\int d^{D}x\left[  \mathbb{\varphi}_{B}\left(  x\right)
\right]  ^{4}%
\end{equation}

where the scalar field $\mathbb{\varphi}_{B}$ is a bare field, $m_{B}$ is the
bare mass, and $\mu$ is an arbitrary mass parameter introduced to keep the
bare coupling constant $g_{B}$ dimensionless. The bare quantities, such as
$\mathbb{\varphi}_{B},$ $m_{B}$ and $g_{B}$, are objects which are useful in
the intermediate steps of the calculations, but they have no physical meaning.
They are just chosen so that they cancel the divergences and leave us with the
desired renormalized quantity \cite{MAG} \cite{WEINB} \cite{QFT3}. The
physical or renormalized scalar field $\mathbb{\varphi}$, mass $m$ and
coupling constant $g$ are defined by \cite{MAG} \cite{WEINB} \cite{QFT3}
\cite{QFT2}%

\begin{align}
\mathbb{\varphi}  &  \mathbb{=}Z^{-\frac{1}{2}}\mathbb{\varphi}_{B}\text{
\ \ , \ \ }m^{2}=m_{B}^{2}+\delta m^{2}\\
g  &  =Z_{g}^{-\frac{1}{2}}g_{B}\text{ \ \ , \ \ }Z_{g}=\frac{\left(
1+B\right)  }{Z^{2}}%
\end{align}

with $Z$ and $\delta m^{2}$ to be chosen so that the propagators of the
renormalized fields have poles in the same position and with the same residues
as the propagators of the free fields in the absence of interactions.

The action may then be written in terms of renormalized quantities, as%

\begin{equation}
\mathcal{S}\left[  \varphi\right]  =\mathcal{S}^{\left(  0\right)  }\left[
\varphi\right]  +\mathcal{S}^{\left(  1\right)  }\left[  \varphi\right]
+\mathcal{S}^{\left(  c\right)  }\left[  \varphi\right]
\end{equation}

where%

\begin{align}
\mathcal{S}^{\left(  0\right)  }\left[  \varphi\right]   &  =-\frac{1}{2}\int
d^{D}xd^{D}y\mathbb{\varphi}\left(  x\right)  \mathcal{D}\left(  x,y\right)
\mathbb{\varphi}\left(  y\right) \\
\mathcal{S}^{\left(  1\right)  }\left[  \varphi\right]   &  =-\frac{1}{4!}%
\mu^{4-D}g\int d^{D}x\left[  \mathbb{\varphi}\left(  x\right)  \right]  ^{4}\\
\mathcal{D}\left(  x,y\right)   &  =\left(  m^{2}-\square\right)  \delta
^{4}\left(  x,y\right)  =\int\frac{d^{4}q}{\left(  2\pi\right)  ^{4}}\left(
m^{2}+q^{2}\right)  e^{iq\left(  x-y\right)  }%
\end{align}

and
\begin{equation}
\mathcal{S}^{\left(  c\right)  }\left[  \varphi\right]  =-\frac{1}{2}\left(
Z-1\right)  \int d^{D}xd^{D}y\mathbb{\varphi}\left(  x\right)  \mathcal{D}%
\left(  x,y\right)  \mathbb{\varphi}\left(  y\right)  +\frac{1}{2}Z\delta
m^{2}\int d^{D}x\mathbb{\varphi}\left(  x\right)  \mathbb{\varphi}\left(
x\right)  -\frac{B}{4!}\mu^{4-D}g\int d^{D}x\left[  \mathbb{\varphi}\left(
x\right)  \right]  ^{4}%
\end{equation}

The renormalized exact propagator $\Delta^{\left(  R\right)  }\left(
x,y\right)  $ is given by

\begin{equation}
-i\Delta^{\left(  R\right)  }\left(  x,y\right)  =\left\langle 0\right\vert
T\left\{  \Phi\left(  x\right)  \Phi\left(  y\right)  \right\}  \left\vert
0\right\rangle =\frac{\int%
{\displaystyle\prod\limits_{x}}
d\mathbb{\varphi}\left(  x\right)  \left[  \mathbb{\varphi}\left(  x\right)
\mathbb{\varphi}\left(  y\right)  \right]  e^{i\mathcal{S}\left[
\varphi\right]  }}{\int%
{\displaystyle\prod\limits_{x}}
d\mathbb{\varphi}\left(  x\right)  e^{i\mathcal{S}\left[  \varphi\right]  }}
\label{eqn a10}%
\end{equation}

We would like to evaluate the path integral for this theory, we now make a
Taylor expansion in powers of g\ %

\begin{equation}
e^{i\mathcal{S}\left[  \varphi\right]  }=e^{i\mathcal{S}^{\left(  0\right)
}\left[  \varphi\right]  +i\mathcal{S}^{\left(  1\right)  }\left[
\varphi\right]  }=e^{i\mathcal{S}^{\left(  0\right)  }\left[  \varphi\right]
}\left(  1+i\mathcal{S}^{\left(  1\right)  }\left[  \varphi\right]
+i\mathcal{S}^{\left(  c\right)  }\left[  \varphi\right]  +...\right)
\end{equation}

Using this Taylor expansion in eq$\left(  \ref{eqn a10}\right)  $ we obtain
the following expressions%

\begin{multline}
\int%
{\displaystyle\prod\limits_{x}}
d\mathbb{\varphi}\left(  x\right)  \left[  \mathbb{\varphi}\left(  x\right)
\mathbb{\varphi}\left(  y\right)  \right]  e^{i\mathcal{S}\left[
\varphi\right]  }=\int%
{\displaystyle\prod\limits_{x}}
d\mathbb{\varphi}\left(  x\right)  \left[  \mathbb{\varphi}\left(  x\right)
\mathbb{\varphi}\left(  y\right)  \right]  e^{i\mathcal{S}^{\left(  0\right)
}\left[  \varphi\right]  }\\
-\frac{i}{4!}\mu^{4-D}g\int d^{D}z\int%
{\displaystyle\prod\limits_{x}}
d\mathbb{\varphi}\left(  x\right)  \left[  \mathbb{\varphi}\left(  x\right)
\mathbb{\varphi}\left(  y\right)  \left[  \varphi\left(  z\right)  \right]
^{4}\right]  e^{i\mathcal{S}^{\left(  0\right)  }\left[  \varphi\right]
}\nonumber\\
-\frac{i}{2}\left(  Z-1\right)  \int d^{D}zd^{D}\tau\mathcal{D}\left(
z,\tau\right)  \int%
{\displaystyle\prod\limits_{x}}
d\mathbb{\varphi}\left(  x\right)  \left[  \mathbb{\varphi}\left(  x\right)
\mathbb{\varphi}\left(  y\right)  \varphi\left(  z\right)  \mathbb{\varphi
}\left(  \tau\right)  \right]  e^{i\mathcal{S}^{\left(  0\right)  }\left[
\varphi\right]  }\nonumber\\
+\frac{i}{2}Z\delta m^{2}\int d^{D}z\int%
{\displaystyle\prod\limits_{x}}
d\mathbb{\varphi}\left(  x\right)  \left[  \mathbb{\varphi}\left(  x\right)
\mathbb{\varphi}\left(  y\right)  \left[  \varphi\left(  z\right)  \right]
^{2}\right]  e^{i\mathcal{S}^{\left(  0\right)  }\left[  \varphi\right]
}\nonumber\\
-\frac{iB}{4!}\mu^{4-D}g\int d^{D}z\int%
{\displaystyle\prod\limits_{x}}
d\mathbb{\varphi}\left(  x\right)  \left[  \mathbb{\varphi}\left(  x\right)
\mathbb{\varphi}\left(  y\right)  \left[  \varphi\left(  z\right)  \right]
^{4}\right]  e^{i\mathcal{S}^{\left(  0\right)  }\left[  \varphi\right]
}\nonumber
\end{multline}

and%

\begin{multline}
\int%
{\displaystyle\prod\limits_{x}}
d\mathbb{\varphi}\left(  x\right)  e^{i\mathcal{S}\left[  \varphi\right]
}=\int%
{\displaystyle\prod\limits_{x}}
d\mathbb{\varphi}\left(  x\right)  e^{i\mathcal{S}^{\left(  0\right)  }\left[
\varphi\right]  }-\frac{i}{4!}\mu^{4-D}g\int d^{D}z\int%
{\displaystyle\prod\limits_{x}}
d\mathbb{\varphi}\left(  x\right)  \left[  \varphi\left(  z\right)  \right]
^{4}e^{i\mathcal{S}^{\left(  0\right)  }\left[  \varphi\right]  }\\
-\frac{i}{2}\left(  Z-1\right)  \int d^{D}zd^{D}\tau\mathcal{D}\left(
z,\tau\right)  \int%
{\displaystyle\prod\limits_{x}}
d\mathbb{\varphi}\left(  x\right)  \left[  \varphi\left(  z\right)
\mathbb{\varphi}\left(  \tau\right)  \right]  e^{i\mathcal{S}^{\left(
0\right)  }\left[  \varphi\right]  }\nonumber\\
+\frac{1}{2}Z\delta m^{2}\int d^{D}z\int%
{\displaystyle\prod\limits_{x}}
d\mathbb{\varphi}\left(  x\right)  \left[  \varphi\left(  z\right)  \right]
^{2}e^{i\mathcal{S}^{\left(  0\right)  }\left[  \varphi\right]  }\nonumber\\
-\frac{B}{4!}\mu^{4-D}g\int d^{D}z\int%
{\displaystyle\prod\limits_{x}}
d\mathbb{\varphi}\left(  x\right)  \left[  \varphi\left(  z\right)  \right]
^{4}e^{i\mathcal{S}^{\left(  0\right)  }\left[  \varphi\right]  }\nonumber
\end{multline}

The functional integrals in eq$\left(  \ref{eqn a10}\right)  $\ can now be
performed with the help of the identity eq$\left(  \ref{eqn a7}\right)  $, the
renormalized exact propagator $\Delta^{\left(  R\right)  }\left(  x,y\right)
$ can be written as%

\begin{equation}
\Delta^{\left(  R\right)  }\left(  x,y\right)  =\int\frac{d^{D}q}{\left(
2\pi\right)  ^{D}}\Delta^{\left(  R\right)  }\left(  q\right)  e^{iq\left(
x-y\right)  }%
\end{equation}

with%

\begin{equation}
\Delta^{\left(  R\right)  }\left(  q\right)  =\Delta\left(  q\right)
+\Delta\left(  q\right)  \left[  \Pi^{\ast}\left(  q^{2}\right)  \right]
\Delta\left(  q\right)  =\frac{1}{q^{2}+m^{2}-\Pi^{\ast}\left(  q^{2}\right)
-i\epsilon}+O\left(  g^{2}\right)
\end{equation}

where $\Pi^{\ast}\left(  q^{2}\right)  $ is the self-energy function%

\begin{equation}
\Pi^{\ast}\left(  q^{2}\right)  =-\left(  Z-1\right)  \left(  m^{2}%
+q^{2}\right)  +Z\delta m^{2}+\Pi_{Loop}^{\ast}\left(  q^{2}\right)
\end{equation}

and%

\begin{equation}
\Pi_{Loop}^{\ast}\left(  q^{2}\right)  =\frac{i}{2}\mu^{4-D}g\int\frac{d^{D}%
p}{\left(  2\pi\right)  ^{4}}\frac{1}{p^{2}+m^{2}-i\epsilon}%
\end{equation}

The condition that $m^{2}$ is the true mass of the particle is that the pole
of the propagator should be at $q^{2}=-m^{2}$, so that \cite{WEINB}%

\begin{equation}
\Pi^{\ast}\left(  -m^{2}\right)  =0
\end{equation}

Also, the condition that the pole of the propagator at $q^{2}=-m^{2}$ should
have a unit residue (like the uncorrected propagator) is that%
\begin{equation}
\left[  \frac{d}{dq^{2}}\Pi^{\ast}\left(  q^{2}\right)  \right]
_{q^{2}=-m^{2}}=0
\end{equation}

These conditions allow us to evaluate $Z$ and $\delta m^{2}$ \cite{WEINB}%

\begin{equation}
Z\delta m^{2}=-\Pi_{Loop}^{\ast}\left(  -m^{2}\right)
\end{equation}

and%

\begin{equation}
Z=1+\left[  \frac{d}{dq^{2}}\Pi_{Loop}^{\ast}\left(  q^{2}\right)  \right]
_{q^{2}=-m^{2}}%
\end{equation}

Direct calculations give the following expression for the loop integral
\cite{QFT1}-\cite{QFT4} \cite{Collins}%

\begin{equation}
\Pi_{Loop}^{\ast}\left(  q^{2}\right)  =\frac{i}{2}\mu^{4-D}g\int\frac{d^{D}%
p}{\left(  2\pi\right)  ^{4}}\frac{1}{p^{2}+m^{2}-i\epsilon}=-g\frac{m^{2}%
}{32\pi^{2}}\left(  \frac{4\pi\mu^{2}}{m^{2}}\right)  ^{2-\frac{D}{2}}%
\Gamma\left(  1-\frac{D}{2}\right)
\end{equation}

The divergence of this integral manifests itself in the pole of the Gamma
function at the physical dimension $D=4$.

In the neighborhood of this dimension we set%
\begin{equation}
D=4-2s
\end{equation}

then%

\begin{equation}
\Pi_{Loop}^{\ast}\left(  q^{2}\right)  =-g\frac{m^{2}}{32\pi^{2}}\left(
\frac{m^{2}}{4\pi\mu^{2}}\right)  ^{-s}\Gamma\left(  s-1\right)
\end{equation}

using the relations \cite{TOMS} \cite{Witt 2} \cite{QFT2}%

\begin{equation}
\Gamma\left(  s-1\right)  =-\left(  \frac{1}{s}-\gamma+1\right)  +O\left(
s\right)
\end{equation}

and%
\begin{equation}
\left(  \frac{m^{2}}{4\pi\mu^{2}}\right)  ^{-s}=1-s\ln\left(  \frac{m^{2}%
}{4\pi\mu^{2}}\right)  +O\left(  s\right)
\end{equation}

we get%

\begin{equation}
\Pi_{Loop}^{\ast}\left(  q^{2}\right)  =g\frac{m^{2}}{32\pi^{2}}\left[
\frac{1}{s}+1-\gamma-\ln\left(  \frac{m^{2}}{4\pi\mu^{2}}\right)  \right]
+O\left(  s\right)
\end{equation}

hence%

\begin{equation}
Z=1+\left[  \frac{d}{dq^{2}}\Pi_{Loop}^{\ast}\left(  q^{2}\right)  \right]
_{q^{2}=-m^{2}}=1+O\left(  g^{2}\right)
\end{equation}

and%

\begin{equation}
Z\delta m^{2}=\delta m^{2}=-\Pi_{Loop}^{\ast}\left(  -m^{2}\right)
=g\frac{m^{2}}{32\pi^{2}}\frac{1}{s}+g\frac{m^{2}}{32\pi^{2}}\left[
1-\gamma-\ln\left(  \frac{m^{2}}{4\pi\mu^{2}}\right)  \right]  +O\left(
g^{2}\right)
\end{equation}

to this order $\Pi^{\ast}\left(  q^{2}\right)  $ is equal to zero%

\begin{equation}
\Pi^{\ast}\left(  q^{2}\right)  =0+O\left(  g^{2}\right)
\end{equation}

\section{Path Integral Derivation of Noncommutative Propagator}

\subsection{Noncommutative Free Propagator}

The noncommutative action eq$\left(  \ref{eqn17}\right)  $\ can be rewritten as%

\begin{equation}
\widehat{\mathcal{S}}\left[  \varphi\right]  =\int d^{4}x\widehat
{\mathfrak{L}}=\widehat{\mathcal{S}}^{\left(  0\right)  }\left[
\varphi\right]  +\widehat{\mathcal{S}}^{\left(  1\right)  }\left[
\varphi\right]  +\widehat{\mathcal{S}}^{\left(  2\right)  }\left[
\varphi\right]  +\widehat{\mathcal{S}}^{\left(  3\right)  }\left[
\varphi\right]  +\widehat{\mathcal{S}}^{\left(  4\right)  }\left[
\varphi\right]  +\widehat{\mathcal{S}}^{\left(  5\right)  }\left[
\varphi\right]
\end{equation}

with%

\begin{gather}
\widehat{\mathcal{S}}^{\left(  0\right)  }\left[  \varphi\right]  =-\frac
{1}{2}\int d^{4}xd^{4}y\mathbb{\varphi}_{a}\left(  x\right)  \widetilde
{\mathcal{D}}_{ab}\left(  x,y\right)  \mathbb{\varphi}_{b}\left(  y\right)
\label{eqn a11}\\
\widehat{\mathcal{S}}^{\left(  1\right)  }\left[  \varphi\right]  =-\frac
{1}{4}g\int d^{4}x\left[  \mathbb{\varphi}_{a}\left(  x\right)
\mathbb{\varphi}_{a}\left(  x\right)  \mathbb{\varphi}_{b}\left(  x\right)
\mathbb{\varphi}_{b}\left(  x\right)  \right] \label{eqn a12}\\
\widehat{\mathcal{S}}^{\left(  2\right)  }\left[  \varphi\right]  =\frac{1}%
{2}g\theta\int d^{4}xd^{4}y\mathcal{K}_{ab}^{\left(  1\right)  }\left(
x,y\right)  \left[  \mathbb{\varphi}_{a}\left(  x\right)  \mathbb{\varphi}%
_{b}\left(  y\right)  \mathbb{\varphi}_{c}\left(  x\right)  \mathbb{\varphi
}_{c}\left(  x\right)  \right] \label{eqn a13}\\
\widehat{\mathcal{S}}^{\left(  3\right)  }\left[  \varphi\right]  =\frac{1}%
{4}g\theta^{2}\int d^{4}xd^{4}yd^{4}\tau\mathcal{K}^{\left(  2\right)
}\left(  x;y,\tau\right)  \left[  \mathbb{\varphi}_{a}\left(  x\right)
\mathbb{\varphi}_{a}\left(  y\right)  \mathbb{\varphi}_{c}\left(  \tau\right)
\mathbb{\varphi}_{c}\left(  \tau\right)  \right] \label{eqn a14}\\
\widehat{\mathcal{S}}^{\left(  4\right)  }\left[  \varphi\right]  =-\frac
{1}{8}g\theta^{2}\int d^{4}xd^{4}yd^{4}\tau\mathcal{K}^{\left(  3\right)
}\left(  x;y,\tau\right)  \left[  \mathbb{\varphi}_{a}\left(  y\right)
\mathbb{\varphi}_{a}\left(  \tau\right)  \mathbb{\varphi}_{c}\left(  x\right)
\mathbb{\varphi}_{c}\left(  x\right)  \right] \label{eqn a15}\\
\widehat{\mathcal{S}}^{\left(  5\right)  }\left[  \varphi\right]  =-\frac
{1}{4}g\theta^{2}\int d^{4}xd^{4}yd^{4}\tau\mathcal{K}^{\left(  3\right)
}\left(  x;y,\tau\right)  \varepsilon_{ad}\varepsilon_{bc}\left[
\mathbb{\varphi}_{a}\left(  y\right)  \mathbb{\varphi}_{d}\left(  x\right)
\mathbb{\varphi}_{b}\left(  \tau\right)  \mathbb{\varphi}_{c}\left(  x\right)
\right]  \label{eqn a16}%
\end{gather}

where $\widetilde{\mathcal{D}}_{ab}\left(  x,y\right)  $ is the symmetric matrix%

\begin{gather}
\widetilde{\mathcal{D}}_{ab}\left(  x,y\right)  =\mathcal{D}_{ab}\left(
x,y\right)  -\frac{1}{4}\theta^{2}\mathcal{M}_{ab}\left(  x,y\right)
-\theta\mathcal{N}_{ab}\left(  x,y\right)  =\widehat{\mathcal{D}}_{ba}\left(
y,x\right) \label{eqn c2}\\
\mathcal{D}_{ab}\left(  x,y\right)  =\delta_{ab}\left(  m^{2}-\square\right)
\delta^{4}\left(  x,y\right)  =\delta_{ab}\int\frac{d^{4}q}{\left(
2\pi\right)  ^{4}}\left(  m^{2}+q^{2}\right)  e^{iq\left(  x-y\right)
}=\mathcal{D}_{ba}\left(  y,x\right) \label{eqn c13}\\
\mathcal{M}_{ab}\left(  x,y\right)  =\delta_{ab}\left(  m^{2}-\square\right)
\left(  m^{2}-\overrightarrow{\nabla}^{2}\right)  \delta^{4}\left(
x,y\right)  =\delta_{ab}\int\frac{d^{4}q}{\left(  2\pi\right)  ^{4}}\left(
m^{2}+q^{2}\right)  \left(  m^{2}+\overrightarrow{q}^{2}\right)  e^{iq\left(
x-y\right)  }\label{eqn c14}\\
\mathcal{M}_{ab}\left(  x,y\right)  =\mathcal{M}_{ba}\left(  y,x\right)
\label{eqn c15}\\
\mathcal{N}_{ab}\left(  x,y\right)  =\left(  m^{2}-\overrightarrow{\nabla}%
^{2}\right)  \varepsilon_{ab}\partial_{t}\delta^{4}\left(  x,y\right)
=\varepsilon_{ab}\int\frac{d^{4}q}{\left(  2\pi\right)  ^{4}}\left(
m^{2}+\overrightarrow{q}^{2}\right)  \left[  iq_{0}\right]  e^{iq\left(
x-y\right)  }=\mathcal{N}_{ba}\left(  y,x\right)  \label{eqn c16}%
\end{gather}

while $\mathcal{K}_{ab}^{\left(  1\right)  }\left(  x,y\right)  $ ,
$\mathcal{K}^{\left(  2\right)  }\left(  x;y,\tau\right)  $\ and
$\mathcal{K}^{\left(  3\right)  }\left(  x;y,\tau\right)  $\ are given by%

\begin{gather}
\mathcal{K}_{ab}^{\left(  1\right)  }\left(  x,y\right)  =\varepsilon
_{ab}\partial_{t}\delta^{4}\left(  x,y\right)  =\int\frac{d^{4}q}{\left(
2\pi\right)  ^{4}}\left[  \varepsilon_{ab}iq_{0}\right]  e^{iq\left(
x-y\right)  }=\mathcal{K}_{ba}^{\left(  1\right)  }\left(  y,x\right) \\
\mathcal{K}^{\left(  2\right)  }\left(  x;y,\tau\right)  =\left[  \delta
^{4}\left(  x,\tau\right)  \left(  m^{2}-\overrightarrow{\nabla}^{2}\right)
\delta^{4}\left(  x,y\right)  -\left(  \overrightarrow{\nabla}\delta
^{4}\left(  x,\tau\right)  \right)  \left(  \overrightarrow{\nabla}\delta
^{4}\left(  x,y\right)  \right)  -\frac{1}{2}\left(  \overrightarrow{\nabla
}^{2}\delta^{4}\left(  x,\tau\right)  \right)  \delta^{4}\left(  x,y\right)
\right] \\
\mathcal{K}^{\left(  2\right)  }\left(  x;y,\tau\right)  =\int\frac{d^{4}%
q}{\left(  2\pi\right)  ^{4}}\frac{d^{4}p}{\left(  2\pi\right)  ^{4}}\left[
m^{2}+\frac{1}{2}\left(  \overrightarrow{p}+\overrightarrow{q}\right)
^{2}+\frac{1}{2}\overrightarrow{p}^{2}\right]  e^{iq\left(  x-\tau\right)
}e^{ip\left(  x-y\right)  }%
\end{gather}

\begin{gather}
\mathcal{K}^{\left(  3\right)  }\left(  x;y,\tau\right)  =\left[  \partial
_{t}\delta^{4}\left(  x,y\right)  \right]  \left[  \partial_{t}\delta
^{4}\left(  x,\tau\right)  \right] \\
\mathcal{K}^{\left(  3\right)  }\left(  x;y,\tau\right)  =-\int\frac{d^{4}%
q}{\left(  2\pi\right)  ^{4}}\frac{d^{4}p}{\left(  2\pi\right)  ^{4}}%
q_{0}p_{0}e^{iq\left(  x-y\right)  }e^{ip\left(  x-\tau\right)  }%
=\mathcal{K}^{\left(  3\right)  }\left(  x;\tau,y\right)
\end{gather}

The free noncommutative propagator $\widetilde{\Delta}_{ab}\left(  x,y\right)
$ is defined as the vacuum expectation value%
\begin{equation}
-i\widetilde{\Delta}_{ab}\left(  x,y\right)  =\left\langle 0\right\vert
T\left\{  \Phi_{a}\left(  x\right)  \Phi_{b}\left(  y\right)  \right\}
\left\vert 0\right\rangle
\end{equation}

where $\left\langle 0\right\vert T\left\{  \mathcal{A}\left[  \Phi\right]
\right\}  \left\vert 0\right\rangle $\ denotes the vacuum expectation value of
the chronological product $T\left\{  \mathcal{A}\left[  \Phi\right]  \right\}
$, $\Phi\left(  x\right)  $ denotes the free quantum field operator
corresponding to $\mathbb{\varphi}\left(  x\right)  $, and $T$ is the
time-ordering operator.

The vacuum expectation value eq$\left(  \ref{eqn a1}\right)  $ can be
expressed as a Feynman functional integral%
\begin{equation}
-i\widetilde{\Delta}_{ab}\left(  x,y\right)  =\left\langle 0\right\vert
T\left\{  \Phi_{a}\left(  x\right)  \Phi_{b}\left(  y\right)  \right\}
\left\vert 0\right\rangle =\frac{\int%
{\displaystyle\prod\limits_{a,x}}
d\mathbb{\varphi}_{a}\left(  x\right)  \left[  \mathbb{\varphi}_{a}\left(
x\right)  \mathbb{\varphi}_{b}\left(  y\right)  \right]  e^{i\widehat
{\mathcal{S}}^{\left(  0\right)  }\left[  \varphi\right]  }}{\int%
{\displaystyle\prod\limits_{a,x}}
d\mathbb{\varphi}_{a}\left(  x\right)  e^{i\widehat{\mathcal{S}}^{\left(
0\right)  }\left[  \varphi\right]  }}%
\end{equation}

The functional integrals in eq$\left(  \ref{eqn a4}\right)  $ can now be
performed with the help of the identity%

\begin{align}
&  \int%
{\displaystyle\prod\limits_{x}}
d\mathbb{\varphi}\left(  x\right)  \left[  \mathbb{\varphi}_{l_{1}}\left(
x_{1}\right)  \mathbb{\varphi}_{_{l_{2}}}\left(  x_{2}\right)
...\mathbb{\varphi}_{_{l_{2N}}}\left(  x_{2N}\right)  \right]  e^{i\widehat
{\mathcal{S}}^{\left(  0\right)  }\left[  \varphi\right]  }\label{eqn b1}\\
&  =\left[  \det\left(  \frac{i\widetilde{\mathcal{D}}}{2\pi}\right)  \right]
^{-\frac{1}{2}}\sum_{\substack{pairings\\of\ \left(  x_{1}l_{1},x_{2}%
l_{2},...,x_{2N}l_{2N}\right)  }}%
{\displaystyle\prod\limits_{pairs\left(  x_{j}l_{j},x_{k}l_{k}\right)  }}
\left[  -i\widetilde{\mathcal{D}}_{l_{j},l_{k}}^{-1}\left(  x_{j}%
,x_{k}\right)  \right] \nonumber
\end{align}

where the sum is over all ways of pairing the indices $\left(  x_{1}%
l_{1},x_{2}l_{2},...,x_{2N}l_{2N}\right)  $, with two pairings being
considered the same if they differ only by the order of the pairs, or by the
order of indices within a pair, and $\widetilde{\mathcal{D}}_{ab}^{-1}\left(
x,y\right)  $ is the inverse of the matrix $\widetilde{\mathcal{D}}%
_{ab}\left(  x,y\right)  .$

Direct calculations give the following expression for the propagator%
\begin{equation}
-i\widetilde{\Delta}_{ab}\left(  x,y\right)  =-i\widetilde{\mathcal{D}}%
_{ab}^{-1}\left(  x,y\right)  \label{eqn b2}%
\end{equation}

to get the inverse of the matrix $\widetilde{\mathcal{D}}_{ab}\left(
x,y\right)  $ we write $\widetilde{\mathcal{D}}_{ab}\left(  x,y\right)  $ as a
Fourier integral%

\begin{equation}
\widetilde{\mathcal{D}}_{ab}\left(  x,y\right)  =\int\frac{d^{4}q}{\left(
2\pi\right)  ^{4}}\widetilde{\mathcal{D}}_{ab}\left(  q\right)  e^{iq\left(
x-y\right)  }%
\end{equation}

where%
\begin{equation}
\widetilde{\mathcal{D}}_{ab}\left(  q\right)  =\left(  m^{2}+q^{2}\right)
\delta_{ab}-\frac{1}{4}\theta^{2}\delta_{ab}\left(  m^{2}+q^{2}\right)
\left(  m^{2}+\overrightarrow{q}^{2}\right)  -i\theta\varepsilon_{ab}\left(
m^{2}+\overrightarrow{q}^{2}\right)  q_{0} \label{eqn c1}%
\end{equation}

Now using these relations and the fact that eq$\left(  \ref{eqn b2}\right)  $
can be rewritten as

\begin{equation}
\int d^{4}y\widetilde{\Delta}_{ab}\left(  x,y\right)  \widetilde{\mathcal{D}%
}_{bc}\left(  y,z\right)  =\delta_{ac}\delta^{4}\left(  x,z\right)
\end{equation}

one can deduce that%

\begin{equation}
\widetilde{\Delta}_{ab}\left(  x,y\right)  =\int\frac{d^{4}q}{\left(
2\pi\right)  ^{4}}\widetilde{\Delta}_{ab}\left(  q\right)  e^{iq\left(
x-y\right)  }%
\end{equation}

with%
\begin{equation}
\widetilde{\Delta}_{ab}\left(  q\right)  =\widetilde{\mathcal{D}}_{ab}%
^{-1}\left(  q\right)  =\frac{1}{m^{2}+q^{2}}\delta_{ab}+\frac{1}{4}\theta
^{2}\delta_{ab}\frac{m^{2}+\overrightarrow{q}^{2}}{m^{2}+q^{2}}+\theta
^{2}\delta_{ab}\frac{\left(  m^{2}+\overrightarrow{q}^{2}\right)  ^{2}%
}{\left(  m^{2}+q^{2}\right)  ^{3}}\left(  q_{0}\right)  ^{2}+i\theta
\varepsilon_{ab}\frac{m^{2}+\overrightarrow{q}^{2}}{\left(  m^{2}%
+q^{2}\right)  ^{2}}q_{0}%
\end{equation}

it is easy to show that the free noncommutative propagator has the following
symmetry properties%

\begin{gather}
\widetilde{\Delta}_{ab}\left(  q\right)  =\widetilde{\mathcal{D}}_{ab}%
^{-1}\left(  q\right)  =\delta_{ab}\widetilde{\Delta}\left(  q\right)
+\varepsilon_{ab}\widetilde{\eta}\left(  q\right)  =\delta_{ab}\widetilde
{\Delta}\left(  q\right)  +\theta\varepsilon_{ab}\overline{\eta}\left(
q\right)  =\widetilde{\Delta}_{ba}\left(  -q\right) \\
\widetilde{\Delta}\left(  q\right)  \ \ =\frac{1}{m^{2}+q^{2}}+\frac{1}%
{4}\theta^{2}\frac{m^{2}+\overrightarrow{q}^{2}}{m^{2}+q^{2}}+\theta^{2}%
\frac{\left(  m^{2}+\overrightarrow{q}^{2}\right)  ^{2}}{\left(  m^{2}%
+q^{2}\right)  ^{3}}\left(  q_{0}\right)  ^{2}=\widetilde{\Delta}\left(
-q\right) \\
\ \widetilde{\eta}\left(  q\right)  =\theta\overline{\eta}\left(  q\right)
=i\theta\frac{m^{2}+\overrightarrow{q}^{2}}{\left(  m^{2}+q^{2}\right)  ^{2}%
}q_{0}=-\ \widetilde{\eta}\left(  -q\right)
\end{gather}

\subsection{Noncommutative Exact Propagator}

The exact noncommutative propagator $\widetilde{\Delta}_{ab}^{\prime}\left(
x,y\right)  $ is given by%
\begin{equation}
-i\widetilde{\Delta}_{ab}^{\prime}\left(  x,y\right)  =\left\langle
0\right\vert T\left\{  \Phi_{a}\left(  x\right)  \Phi_{b}\left(  y\right)
\right\}  \left\vert 0\right\rangle =\frac{\int%
{\displaystyle\prod\limits_{a,x}}
d\mathbb{\varphi}_{a}\left(  x\right)  \left[  \mathbb{\varphi}_{a}\left(
x\right)  \mathbb{\varphi}_{b}\left(  y\right)  \right]  e^{i\widehat
{\mathcal{S}}\left[  \varphi\right]  }}{\int%
{\displaystyle\prod\limits_{a,x}}
d\mathbb{\varphi}_{a}\left(  x\right)  e^{i\widehat{\mathcal{S}}\left[
\varphi\right]  }} \label{eqn b3}%
\end{equation}

In order to evaluate the functional integral eq$\left(  \ref{eqn b3}\right)
$, we make a Taylor expansion in powers of g\ %

\begin{equation}
e^{i\widehat{\mathcal{S}}\left[  \varphi\right]  }=\left[  1+i%
{\displaystyle\sum\limits_{k=1}^{5}}
\widehat{\mathcal{S}}^{\left(  k\right)  }\left[  \varphi\right]  \right]
e^{i\widehat{\mathcal{S}}^{\left(  0\right)  }\left[  \varphi\right]  }
\label{eqn b4}%
\end{equation}

Using this Taylor expansion in eq$\left(  \ref{eqn b3}\right)  $ and expanding
the denominator by the binomial theorem, we obtain the following expression
for the exact noncommutative propagator%

\begin{equation}
-i\widetilde{\Delta}_{ab}^{\prime}\left(  x,y\right)  =-i\widetilde{\Delta
}_{ab}\left(  x,y\right)  -\frac{i}{4}g\mathcal{I}_{ab}^{\left(  1\right)
}\left(  x,y\right)  +\frac{i}{2}\theta g\mathcal{I}_{ab}^{\left(  2\right)
}\left(  x,y\right)  +\frac{i}{4}g\theta^{2}\left[  \mathcal{I}_{ab}^{\left(
3\right)  }\left(  x,y\right)  -\frac{1}{2}\mathcal{I}_{ab}^{\left(  4\right)
}\left(  x,y\right)  -\mathcal{I}_{ab}^{\left(  5\right)  }\left(  x,y\right)
\right]
\end{equation}

with%

\begin{gather*}
\mathcal{I}_{ab}^{\left(  1\right)  }\left(  x,y\right)  =\int d^{4}%
z\int_{\left(  c\right)  }%
{\displaystyle\prod\limits_{a,x}}
d\mathbb{\varphi}_{a}\left(  x\right)  \left[  \mathbb{\varphi}_{a}\left(
x\right)  \mathbb{\varphi}_{b}\left(  y\right)  \right]  \left[
\mathbb{\varphi}_{c}\left(  z\right)  \mathbb{\varphi}_{c}\left(  z\right)
\mathbb{\varphi}_{d}\left(  z\right)  \mathbb{\varphi}_{d}\left(  z\right)
\right]  e^{i\widehat{\mathcal{S}}^{\left(  0\right)  }\left[  \varphi\right]
}\\
\mathcal{I}_{ab}^{\left(  2\right)  }\left(  x,y\right)  =\int d^{4}\tau
d^{4}z\mathcal{K}_{cd}^{\left(  1\right)  }\left(  \tau,z\right)
\int_{\left(  c\right)  }%
{\displaystyle\prod\limits_{a,x}}
d\mathbb{\varphi}_{a}\left(  x\right)  \left[  \mathbb{\varphi}_{a}\left(
x\right)  \mathbb{\varphi}_{b}\left(  y\right)  \right]  \left[
\mathbb{\varphi}_{c}\left(  \tau\right)  \mathbb{\varphi}_{d}\left(  z\right)
\mathbb{\varphi}_{m}\left(  \tau\right)  \mathbb{\varphi}_{m}\left(
\tau\right)  \right]  e^{i\widehat{\mathcal{S}}^{\left(  0\right)  }\left[
\varphi\right]  }\\
\mathcal{I}_{ab}^{\left(  3\right)  }\left(  x,y\right)  =\int d^{4}z^{\prime
}d^{4}zd^{4}\tau\mathcal{K}^{\left(  2\right)  }\left(  z^{\prime}%
;z,\tau\right)  \int_{\left(  c\right)  }%
{\displaystyle\prod\limits_{a,x}}
d\mathbb{\varphi}_{a}\left(  x\right)  \left[  \mathbb{\varphi}_{a}\left(
x\right)  \mathbb{\varphi}_{b}\left(  y\right)  \right]  \left[
\mathbb{\varphi}_{c}\left(  z^{\prime}\right)  \mathbb{\varphi}_{c}\left(
z\right)  \mathbb{\varphi}_{d}\left(  \tau\right)  \mathbb{\varphi}_{d}\left(
\tau\right)  \right]  e^{i\widehat{\mathcal{S}}^{\left(  0\right)  }\left[
\varphi\right]  }\\
\mathcal{I}_{ab}^{\left(  4\right)  }\left(  x,y\right)  =\int d^{4}z^{\prime
}d^{4}zd^{4}\tau\mathcal{K}^{\left(  3\right)  }\left(  z^{\prime}%
;z,\tau\right)  \int_{\left(  c\right)  }%
{\displaystyle\prod\limits_{a,x}}
d\mathbb{\varphi}_{a}\left(  x\right)  \left[  \mathbb{\varphi}_{a}\left(
x\right)  \mathbb{\varphi}_{b}\left(  y\right)  \right]  \left[
\mathbb{\varphi}_{c}\left(  z\right)  \mathbb{\varphi}_{c}\left(  \tau\right)
\mathbb{\varphi}_{d}\left(  z^{\prime}\right)  \mathbb{\varphi}_{d}\left(
z^{\prime}\right)  \right]  e^{i\widehat{\mathcal{S}}^{\left(  0\right)
}\left[  \varphi\right]  }%
\end{gather*}

and%
\begin{multline*}
\mathcal{I}_{ab}^{\left(  5\right)  }\left(  x,y\right)  =\varepsilon
_{mc}\varepsilon_{ld}\int d^{4}z^{\prime}d^{4}zd^{4}\tau\mathcal{K}^{\left(
3\right)  }\left(  z^{\prime};z,\tau\right)  \times\\
\int_{\left(  c\right)  }%
{\displaystyle\prod\limits_{a,x}}
d\mathbb{\varphi}_{a}\left(  x\right)  \left[  \mathbb{\varphi}_{a}\left(
x\right)  \mathbb{\varphi}_{b}\left(  y\right)  \right]  \left[
\mathbb{\varphi}_{m}\left(  \tau\right)  \mathbb{\varphi}_{c}\left(
z^{\prime}\right)  \mathbb{\varphi}_{l}\left(  z\right)  \mathbb{\varphi}%
_{d}\left(  z^{\prime}\right)  \right]  e^{i\widehat{\mathcal{S}}^{\left(
0\right)  }\left[  \varphi\right]  }%
\end{multline*}

where the subscript $\left(  c\right)  $ being added to the functional
integral $\int_{\left(  c\right)  }%
{\displaystyle\prod\limits_{a,x}}
d\mathbb{\varphi}_{a}\left(  x\right)  ...$ to remind us that $\mathcal{I}%
_{ab}^{\left(  k\right)  }\left(  x,y\right)  ,$ $k=1,2,3,4,5,$ are connected
Green's functions ( connected Green's functions are obtained by disregarding
all terms which factorize into two or more functions with no overlapping arguments).

The connected Green's functions $\mathcal{I}_{ab}^{\left(  k\right)  }\left(
x,y\right)  ,$ $k=1,2,3,4,5,$ can now be calculated with the help of the
identity eq$\left(  \ref{eqn b1}\right)  $, direct but lengthy calculations
lead to the following expressions for the connected Green's functions%

\begin{equation}
\mathcal{I}_{ab}^{\left(  1\right)  }\left(  x,y\right)  =16i\int\frac{d^{4}%
q}{\left(  2\pi\right)  ^{4}}\widetilde{\Delta}_{ac}\left(  q\right)
\widetilde{\Delta}_{cb}\left(  q\right)  e^{iq\left(  x-y\right)  }\int
\frac{d^{4}p}{\left(  2\pi\right)  ^{4}}\widetilde{\Delta}\left(  p\right)
\label{eqn 37}%
\end{equation}

\begin{align}
\mathcal{I}_{ab}^{\left(  2\right)  }\left(  x,y\right)   &  =-8\int
\frac{d^{4}q}{\left(  2\pi\right)  ^{4}}\left[  \widetilde{\Delta}_{ac}\left(
q\right)  \varepsilon_{cd}\widetilde{\Delta}_{db}\left(  q\right)  \right]
q_{0}e^{iq\left(  x-y\right)  }\int\frac{d^{4}p}{\left(  2\pi\right)  ^{4}%
}\widetilde{\Delta}\left(  p\right) \label{eqn 37-b}\\
&  +8\int\frac{d^{4}q}{\left(  2\pi\right)  ^{4}}\left[  \widetilde{\Delta
}_{ac}\left(  q\right)  \widetilde{\Delta}_{cb}\left(  q\right)  \right]
e^{iq\left(  x-y\right)  }\int\frac{d^{4}p}{\left(  2\pi\right)  ^{4}}%
p_{0}\widetilde{\eta}\left(  p\right) \nonumber
\end{align}

\begin{align}
\mathcal{I}_{ab}^{\left(  3\right)  }\left(  x,y\right)   &  =8i\int
\frac{d^{4}q}{\left(  2\pi\right)  ^{4}}\left[  \widetilde{\Delta}_{ac}\left(
q\right)  \widetilde{\Delta}_{cb}\left(  q\right)  \right]  \left(
m^{2}+\overrightarrow{q}^{2}\right)  e^{iq\left(  x-y\right)  }\int\frac
{d^{4}p}{\left(  2\pi\right)  ^{4}}\widetilde{\Delta}\left(  p\right)
\label{eqn 37-c}\\
&  +8i\int\frac{d^{4}q}{\left(  2\pi\right)  ^{4}}\left[  \widetilde{\Delta
}_{ac}\left(  q\right)  \widetilde{\Delta}_{cb}\left(  q\right)  \right]
e^{iq\left(  x-y\right)  }\int\frac{d^{4}p}{\left(  2\pi\right)  ^{4}}\left(
m^{2}+\overrightarrow{p}^{2}\right)  \widetilde{\Delta}\left(  p\right)
\nonumber
\end{align}

\begin{align}
\mathcal{I}_{ab}^{\left(  4\right)  }\left(  x,y\right)   &  =4i\int
\frac{d^{4}q}{\left(  2\pi\right)  ^{4}}\left[  \widetilde{\Delta}_{ac}\left(
q\right)  \widetilde{\Delta}_{cb}\left(  q\right)  \right]  \left(
q_{0}\right)  ^{2}e^{iq\left(  x-y\right)  }\int\frac{d^{4}p}{\left(
2\pi\right)  ^{4}}\widetilde{\Delta}\left(  p\right) \label{eqn 37-d}\\
&  +4i\int\frac{d^{4}q}{\left(  2\pi\right)  ^{4}}\left[  \widetilde{\Delta
}_{ac}\left(  q\right)  \widetilde{\Delta}_{cb}\left(  q\right)  \right]
e^{iq\left(  x-y\right)  }\int\frac{d^{4}p}{\left(  2\pi\right)  ^{4}}\left(
p_{0}\right)  ^{2}\widetilde{\Delta}\left(  p\right) \nonumber
\end{align}

\begin{align}
\mathcal{I}_{ab}^{\left(  5\right)  }\left(  x,y\right)   &  =2i\int
\frac{d^{4}q}{\left(  2\pi\right)  ^{4}}\left[  \widetilde{\Delta}_{ac}\left(
q\right)  \widetilde{\Delta}_{cb}\left(  q\right)  \right]  \left(
q_{0}\right)  ^{2}e^{iq\left(  x-y\right)  }\int\frac{d^{4}p}{\left(
2\pi\right)  ^{4}}\widetilde{\Delta}\left(  p\right) \label{eqn 37-e}\\
&  +2i\int\frac{d^{4}q}{\left(  2\pi\right)  ^{4}}\left[  \widetilde{\Delta
}_{ac}\left(  q\right)  \widetilde{\Delta}_{cb}\left(  q\right)  \right]
e^{iq\left(  x-y\right)  }\int\frac{d^{4}p}{\left(  2\pi\right)  ^{4}}\left(
p_{0}\right)  ^{2}\widetilde{\Delta}\left(  p\right) \nonumber
\end{align}

Now, from eqs$\left(  \ref{eqn 37}\right)  -\left(  \ref{eqn 37-e}\right)  $
one can easily show that%

\begin{equation}
\widetilde{\Delta}_{ab}^{\prime}\left(  x,y\right)  =\int\frac{d^{4}q}{\left(
2\pi\right)  ^{4}}\widetilde{\Delta}_{ab}^{\prime}\left(  q\right)
e^{iq\left(  x-y\right)  } \label{eqn b8}%
\end{equation}

with%

\begin{equation}
-i\widetilde{\Delta}_{ab}^{\prime}\left(  q\right)  =-i\widetilde{\Delta}%
_{ab}\left(  q\right)  +\left[  -i\widetilde{\Delta}_{ac}\left(  q\right)
\right]  \left[  i\Pi_{\left(  Loop\right)  cd}^{\ast}\left(  q\right)
\right]  \left[  -i\widetilde{\Delta}_{db}\left(  q\right)  \right]
\label{eqn b5}%
\end{equation}

where%
\begin{align}
\Pi_{\left(  Loop\right)  ab}^{\ast}\left(  q\right)   &  =4ig\delta
_{ab}\left[  \int\frac{d^{4}p}{\left(  2\pi\right)  ^{4}}\widetilde{\Delta
}\left(  p\right)  \right]  +4\theta g\int\frac{d^{4}p}{\left(  2\pi\right)
^{4}}\left[  q_{0}\varepsilon_{ab}\widetilde{\Delta}\left(  p\right)
-\delta_{ab}p_{0}\widetilde{\eta}\left(  p\right)  \right] \label{eqn b7}\\
&  +i\theta^{2}g\delta_{ab}\int\frac{d^{4}p}{\left(  2\pi\right)  ^{4}}\left[
-2\left(  m^{2}+\overrightarrow{q}^{2}\right)  -2\left(  m^{2}+\overrightarrow
{p}^{2}\right)  +\left(  q_{0}\right)  ^{2}+\left(  p_{0}\right)  ^{2}\right]
\widetilde{\Delta}\left(  p\right) \nonumber
\end{align}

\section{Renormalized Noncommutative Exact Propagator}

\subsection{Renormalized Noncommutative Action}

In order to renormalize the noncommutative complex scalar field we proceed as
in $\varphi^{4}-$ theory of a real self-interacting scalar field, that is we
split the noncommutative action $\widehat{\mathcal{S}}\left[  \varphi
_{B}\right]  ,$ expressed in terms of bare couplings $m_{B},g_{B}$ and bare
fields $\varphi_{B},$ in a part depending on the renormalized parameters $m,g$
and field $\varphi$, and in a counter term part \cite{QFT1}-\cite{QFT4}
\cite{WEINB} \cite{Collins} \cite{MAG}%

\begin{equation}
\widehat{\mathcal{S}}\left[  \varphi_{B}\right]  =\int d^{4}x\widehat
{\mathfrak{L}}=\widehat{\mathcal{S}}_{B}^{\left(  0\right)  }\left[
\varphi_{B}\right]  +\widehat{\mathcal{S}}_{B}^{\left(  1\right)  }\left[
\varphi_{B}\right]  +\widehat{\mathcal{S}}_{B}^{\left(  2\right)  }\left[
\varphi_{B}\right]  +\widehat{\mathcal{S}}_{B}^{\left(  3\right)  }\left[
\varphi_{B}\right]  +\widehat{\mathcal{S}}_{B}^{\left(  4\right)  }\left[
\varphi_{B}\right]  +\widehat{\mathcal{S}}_{B}^{\left(  5\right)  }\left[
\varphi_{B}\right]
\end{equation}

The bare quantities, such as $\mathbb{\varphi}_{B},$ $m_{B},$ and $g_{B}$, are
just chosen so that they cancel the divergences and leave us with the desired
renormalized quantity. The physical or renormalized scalar field
$\mathbb{\varphi}$, mass $m$ and coupling constant $g$ are defined by%

\begin{align}
\mathbb{\varphi}  &  \mathbb{=}Z^{-\frac{1}{2}}\mathbb{\varphi}_{B}\text{
\ \ , \ \ }m^{2}=m_{B}^{2}+\delta m^{2}\\
g  &  =Z_{g}^{-\frac{1}{2}}g_{B}\text{ \ \ , \ \ }Z_{g}=\frac{\left(
1+B\right)  }{Z^{2}}%
\end{align}

with $Z$ and $\delta m^{2}$ to be chosen so that the propagators of the
renormalized fields have poles in the same position and with the same residues
as the propagators of the free fields in the absence of interactions. In this
paper we are interested only to the propagator and its renormalization up to
first order in the parameter $g$, so we can assume that
\begin{align}
\mathbb{\varphi}  &  \mathbb{=\varphi}_{B}\text{ \ \ , \ }g=g_{B}\text{ \ \ ,
\ \ }Z=Z_{g}\text{\ }=1\\
m^{2}  &  =m_{B}^{2}+\delta m^{2}%
\end{align}

The noncommutative action $\widehat{\mathcal{S}}_{B}^{\left(  0\right)
}\left[  \varphi_{B}\right]  $ may then be written in terms of renormalized
quantities, as%

\begin{equation}
\widehat{\mathcal{S}}_{B}^{\left(  0\right)  }\left[  \varphi_{B}\right]
=-\frac{1}{2}\int d^{4}xd^{4}y\mathbb{\varphi}_{a}\left(  x\right)
\widetilde{\mathcal{D}}_{ab}^{\left(  B\right)  }\left(  x,y\right)
\mathbb{\varphi}_{b}\left(  y\right)
\end{equation}

where%

\begin{equation}
\widetilde{\mathcal{D}}_{ab}^{\left(  B\right)  }\left(  x,y\right)
=\mathcal{D}_{ab}^{\left(  B\right)  }\left(  x,y\right)  -\frac{1}{4}%
\theta^{2}\mathcal{M}_{ab}^{\left(  B\right)  }\left(  x,y\right)
-\theta\mathcal{N}_{ab}^{\left(  B\right)  }\left(  x,y\right)
\end{equation}

using the expressions of $\mathcal{D}_{ab}^{\left(  B\right)  }\left(
x,y\right)  ,$ $\mathcal{M}_{ab}^{\left(  B\right)  }\left(  x,y\right)  $ and
$\mathcal{N}_{ab}^{\left(  B\right)  }\left(  x,y\right)  $%

\begin{align}
\mathcal{D}_{ab}^{\left(  B\right)  }\left(  x,y\right)   &  =\mathcal{D}%
_{ab}^{\left(  R\right)  }\left(  x,y\right)  -\delta m^{2}\delta^{4}\left(
x,y\right)  \delta_{ab}\\
\mathcal{M}_{ab}^{\left(  B\right)  }\left(  x,y\right)   &  =\mathcal{M}%
_{ab}^{\left(  R\right)  }\left(  x,y\right)  -\delta m^{2}\left(
m^{2}-\square\right)  \delta^{4}\left(  x,y\right)  \delta_{ab}-\delta
m^{2}\left(  m^{2}-\overrightarrow{\nabla}^{2}\right)  \delta^{4}\left(
x,y\right)  \delta_{ab}\\
\mathcal{N}_{ab}^{\left(  B\right)  }\left(  x,y\right)   &  =\mathcal{N}%
_{ab}^{\left(  R\right)  }\left(  x,y\right)  -\delta m^{2}\partial_{t}%
\delta^{4}\left(  x,y\right)  \varepsilon_{ab}%
\end{align}

one can write $\widetilde{\mathcal{D}}_{ab}^{\left(  B\right)  }\left(
x,y\right)  $ as%

\begin{multline}
\widetilde{\mathcal{D}}_{ab}^{\left(  B\right)  }\left(  x,y\right)
=\widetilde{\mathcal{D}}_{ab}^{\left(  R\right)  }\left(  x,y\right)  -\delta
m^{2}\delta^{4}\left(  x,y\right)  \delta_{ab}+\frac{1}{4}\theta^{2}\delta
m^{2}\left[  \left(  m^{2}-\square\right)  \delta^{4}\left(  x,y\right)
+\left(  m^{2}-\overrightarrow{\nabla}^{2}\right)  \delta^{4}\left(
x,y\right)  \right]  \delta_{ab}\\
+\theta\delta m^{2}\partial_{t}\delta^{4}\left(  x,y\right)  \varepsilon
_{ab}\nonumber
\end{multline}

where $\widetilde{\mathcal{D}}_{ab}^{\left(  R\right)  }\left(  x,y\right)  $
is given by eq$\left(  \ref{eqn c2}\right)  $ \
\begin{gather}
\widetilde{\mathcal{D}}_{ab}^{\left(  R\right)  }\left(  x,y\right)
=\mathcal{D}_{ab}^{\left(  R\right)  }\left(  x,y\right)  -\frac{1}{4}%
\theta^{2}\mathcal{M}_{ab}^{\left(  R\right)  }\left(  x,y\right)
-\theta\mathcal{N}_{ab}^{\left(  R\right)  }\left(  x,y\right) \\
\mathcal{D}_{ab}^{\left(  R\right)  }\left(  x,y\right)  =\delta_{ab}\left(
m^{2}-\square\right)  \delta^{4}\left(  x,y\right)  =\delta_{ab}\int
\frac{d^{4}q}{\left(  2\pi\right)  ^{4}}\left(  m^{2}+q^{2}\right)
e^{iq\left(  x-y\right)  }\\
\mathcal{M}_{ab}^{\left(  R\right)  }\left(  x,y\right)  =\delta_{ab}\left(
m^{2}-\square\right)  \left(  m^{2}-\overrightarrow{\nabla}^{2}\right)
\delta^{4}\left(  x,y\right)  =\delta_{ab}\int\frac{d^{4}q}{\left(
2\pi\right)  ^{4}}\left(  m^{2}+q^{2}\right)  \left(  m^{2}+\overrightarrow
{q}^{2}\right)  e^{iq\left(  x-y\right)  }\\
\mathcal{N}_{ab}^{\left(  R\right)  }\left(  x,y\right)  =\left(
m^{2}-\overrightarrow{\nabla}^{2}\right)  \varepsilon_{ab}\partial_{t}%
\delta^{4}\left(  x,y\right)  =\varepsilon_{ab}\int\frac{d^{4}q}{\left(
2\pi\right)  ^{4}}\left(  m^{2}+\overrightarrow{q}^{2}\right)  \left[
iq_{0}\right]  e^{iq\left(  x-y\right)  }=\mathcal{N}_{ba}\left(  y,x\right)
\end{gather}

hence $\widehat{\mathcal{S}}_{B}^{\left(  0\right)  }\left[  \varphi\right]  $
can be rewritten as%

\begin{equation}
\widehat{\mathcal{S}}_{B}^{\left(  0\right)  }\left[  \varphi\right]
=-\frac{1}{2}\int d^{4}xd^{4}y\mathbb{\varphi}_{a}\left(  x\right)
\widetilde{\mathcal{D}}_{ab}^{\left(  R\right)  }\left(  x,y\right)
\mathbb{\varphi}_{b}\left(  y\right)  +\Delta\widehat{\mathcal{S}}^{\left(
0\right)  }\left[  \varphi\right]  \label{eqn c3}%
\end{equation}

where the counter term part $\Delta\widehat{\mathcal{S}}^{\left(  0\right)
}\left[  \varphi\right]  $ is given by%

\begin{multline}
\Delta\widehat{\mathcal{S}}^{\left(  0\right)  }\left[  \varphi\right]
=\frac{1}{2}\delta m^{2}\int d^{4}xd^{4}y\mathbb{\varphi}_{a}\left(  x\right)
\left[  \delta^{4}\left(  x,y\right)  \delta_{ab}\right]  \mathbb{\varphi}%
_{b}\left(  y\right)  -\frac{1}{2}\theta\delta m^{2}\int d^{4}xd^{4}%
y\mathbb{\varphi}_{a}\left(  x\right)  \left[  \partial_{t}\delta^{4}\left(
x,y\right)  \varepsilon_{ab}\right]  \mathbb{\varphi}_{b}\left(  y\right) \\
-\frac{1}{8}\theta^{2}\delta m^{2}\int d^{4}xd^{4}y\mathbb{\varphi}_{a}\left(
x\right)  \left[  \left(  m^{2}-\square\right)  \delta^{4}\left(  x,y\right)
\right]  \delta_{ab}\mathbb{\varphi}_{b}\left(  y\right) \nonumber\\
-\frac{1}{8}\theta^{2}\delta m^{2}\int d^{4}xd^{4}y\mathbb{\varphi}_{a}\left(
x\right)  \left[  \left(  m^{2}-\overrightarrow{\nabla}^{2}\right)  \delta
^{4}\left(  x,y\right)  \right]  \delta_{ab}\mathbb{\varphi}_{b}\left(
y\right) \nonumber
\end{multline}

The interaction part of the noncommutative action $\widehat{\mathcal{S}%
}\left[  \varphi_{B}\right]  $ may then be written in terms of renormalized
quantities as%

\begin{equation}
\widehat{\mathcal{S}}_{B}^{\left(  k\right)  }\left[  \varphi_{B}\right]
=\widehat{\mathcal{S}}_{R}^{\left(  k\right)  }\left[  \varphi\right]
+O\left(  g^{2}\right)  \text{ \ \ \ }k=1,2,3,4,5 \label{eqn c4}%
\end{equation}

where $\widehat{\mathcal{S}}_{R}^{\left(  k\right)  }\left[  \varphi\right]  $
are given by eq$\left(  \ref{eqn12}\right)  -\left(  \ref{eqn16}\right)  $

Now eq$\left(  \ref{eqn c3}\right)  $ and eq$\left(  \ref{eqn c4}\right)  $
lead to the following expression for the noncommutative action $\widehat
{\mathcal{S}}\left[  \varphi_{B}\right]  $%

\begin{equation}
\widehat{\mathcal{S}}\left[  \varphi_{B}\right]  =\widehat{\mathcal{S}}%
_{R}^{\left(  0\right)  }\left[  \varphi\right]  +\Delta\widehat{\mathcal{S}%
}^{\left(  0\right)  }\left[  \varphi\right]  +\widehat{\mathcal{S}}%
_{R}^{\left(  1\right)  }\left[  \varphi\right]  +\widehat{\mathcal{S}}%
_{R}^{\left(  2\right)  }\left[  \varphi\right]  +\widehat{\mathcal{S}}%
_{R}^{\left(  3\right)  }\left[  \varphi\right]  +\widehat{\mathcal{S}}%
_{R}^{\left(  4\right)  }\left[  \varphi\right]  +\widehat{\mathcal{S}}%
_{R}^{\left(  5\right)  }\left[  \varphi\right]  \label{eqn c5}%
\end{equation}

The renormalized noncommutative exact propagator $\widetilde{\Delta}%
_{ab}^{\left(  R\right)  }\left(  x,y\right)  $ is given by%

\begin{equation}
-i\widetilde{\Delta}_{ab}^{\left(  R\right)  }\left(  x,y\right)
=\left\langle 0\right\vert T\left\{  \Phi_{a}\left(  x\right)  \Phi_{b}\left(
y\right)  \right\}  \left\vert 0\right\rangle =\frac{\int%
{\displaystyle\prod\limits_{a,x}}
d\mathbb{\varphi}_{a}\left(  x\right)  \left[  \mathbb{\varphi}_{a}\left(
x\right)  \mathbb{\varphi}_{b}\left(  y\right)  \right]  e^{i\widehat
{\mathcal{S}}_{B}\left[  \varphi\right]  }}{\int%
{\displaystyle\prod\limits_{a,x}}
d\mathbb{\varphi}_{a}\left(  x\right)  e^{i\widehat{\mathcal{S}}_{B}\left[
\varphi\right]  }}%
\end{equation}

making a Taylor expansion in powers of g, we get%

\begin{equation}
-i\widetilde{\Delta}_{ab}^{\left(  R\right)  }\left(  x,y\right)  =\frac{\int%
{\displaystyle\prod\limits_{a,x}}
d\mathbb{\varphi}_{a}\left(  x\right)  \left[  \mathbb{\varphi}_{a}\left(
x\right)  \mathbb{\varphi}_{b}\left(  y\right)  \right]  \left[
1+i\Delta\widehat{\mathcal{S}}^{\left(  0\right)  }\left[  \varphi\right]  +i%
{\displaystyle\sum\limits_{k=1}^{5}}
\widehat{\mathcal{S}}_{R}^{\left(  k\right)  }\left[  \varphi\right]  \left[
\varphi\right]  \right]  e^{i\widehat{\mathcal{S}}_{R}^{\left(  0\right)
}\left[  \varphi\right]  }}{\int%
{\displaystyle\prod\limits_{a,x}}
d\mathbb{\varphi}_{a}\left(  x\right)  \left[  1+i\Delta\widehat{\mathcal{S}%
}^{\left(  0\right)  }\left[  \varphi\right]  +i%
{\displaystyle\sum\limits_{k=1}^{5}}
\widehat{\mathcal{S}}_{R}^{\left(  k\right)  }\left[  \varphi\right]  \left[
\varphi\right]  \right]  e^{i\widehat{\mathcal{S}}_{R}^{\left(  0\right)
}\left[  \varphi\right]  }} \label{eqn c6}%
\end{equation}

The functional integrals in eq$\left(  \ref{eqn c6}\right)  $\ can now be
performed with the help of the identity eq$\left(  \ref{eqn a7}\right)  $, the
renormalized noncommutative exact propagator $\widetilde{\Delta}_{ab}^{\left(
R\right)  }\left(  x,y\right)  $ can be written as%

\begin{multline}
-i\widetilde{\Delta}_{ab}^{\left(  R\right)  }\left(  x,y\right)
=-i\widetilde{\Delta}_{ab}^{\prime}\left(  x,y\right)  -i\delta m^{2}\int
dzdz^{\prime}\delta^{4}\left(  z,z^{\prime}\right)  \widetilde{\Delta}%
_{ac}\left(  x,z\right)  \widetilde{\Delta}_{cb}\left(  z^{\prime},y\right) \\
+i\theta\delta m^{2}\int dzdz^{\prime}\widetilde{\Delta}_{ac}\left(
x,z\right)  \varepsilon_{cd}\widetilde{\Delta}_{db}\left(  z^{\prime
},y\right)  \partial_{t}\delta^{4}\left(  z,z^{\prime}\right) \nonumber\\
+\frac{i}{4}\theta^{2}\delta m^{2}\int dzdz^{\prime}\widetilde{\Delta}%
_{ac}\left(  x,z\right)  \widetilde{\Delta}_{cb}\left(  z^{\prime},y\right)
\left(  m^{2}-\square\right)  \delta^{4}\left(  z,z^{\prime}\right)
\nonumber\\
+\frac{i}{4}\theta^{2}\delta m^{2}\int dzdz^{\prime}\widetilde{\Delta}%
_{cb}\left(  z^{\prime},y\right)  \left(  m^{2}-\overrightarrow{\nabla}%
^{2}\right)  \widetilde{\Delta}_{ac}\left(  x,z\right) \nonumber
\end{multline}

where $-i\widetilde{\Delta}_{ab}^{\prime}\left(  x,y\right)  $ is given by
eqs$\left(  \ref{eqn b8}\right)  -\left(  \ref{eqn b7}\right)  .$

If we write $\widetilde{\Delta}_{ab}^{\left(  R\right)  }\left(  x,y\right)  $
as a Fourier integral%

\begin{equation}
\widetilde{\Delta}_{ab}^{\left(  R\right)  }\left(  x,y\right)  =\int
\frac{d^{4}q}{\left(  2\pi\right)  ^{4}}\widetilde{\Delta}_{ab}^{\left(
R\right)  }\left(  q\right)  e^{iq\left(  x-y\right)  }%
\end{equation}

then%

\begin{gather}
\widetilde{\Delta}_{ab}^{\left(  R\right)  }\left(  q\right)  =\widetilde
{\Delta}_{ab}^{\prime}\left(  q\right)  +\delta m^{2}\widetilde{\Delta}%
_{ac}\left(  q\right)  \widetilde{\Delta}_{cb}\left(  q\right)  -i\theta\delta
m^{2}q_{0}\widetilde{\Delta}_{ac}\left(  q\right)  \varepsilon_{cd}%
\widetilde{\Delta}_{db}\left(  q\right) \label{eqn c7}\\
\ \ \ \ \ \ \ \ \ \ \ \ \ \ \ \ \ \ \ \ \ \ \ \ \ \ \ \ \ -\frac{1}{4}%
\theta^{2}\delta m^{2}\left(  m^{2}+q^{2}\right)  \widetilde{\Delta}%
_{ac}\left(  q\right)  \widetilde{\Delta}_{cb}\left(  q\right)  -\frac{1}%
{4}\theta^{2}\delta m^{2}\left(  m^{2}+\overrightarrow{q}^{2}\right)
\widetilde{\Delta}_{ac}\left(  q\right)  \widetilde{\Delta}_{cb}\left(
q\right) \nonumber
\end{gather}

where%

\begin{equation}
\widetilde{\Delta}_{ab}^{\prime}\left(  q\right)  =\widetilde{\Delta}%
_{ab}\left(  q\right)  +\widetilde{\Delta}_{ac}\left(  q\right)  \Pi_{\left(
Loop\right)  cd}^{\ast}\left(  q\right)  \widetilde{\Delta}_{db}\left(
q\right)  = \label{eqn c8}%
\end{equation}

and%
\begin{align}
\Pi_{\left(  Loop\right)  ab}^{\ast}\left(  q\right)   &  =4ig\delta
_{ab}\left[  \int\frac{d^{4}p}{\left(  2\pi\right)  ^{4}}\widetilde{\Delta
}\left(  p\right)  \right]  +4\theta g\varepsilon_{ab}\left[  \int\frac
{d^{4}p}{\left(  2\pi\right)  ^{4}}\widetilde{\Delta}\left(  p\right)
\right]  q_{0}-4\theta^{2}g\delta_{ab}\left[  \int\frac{d^{4}p}{\left(
2\pi\right)  ^{4}}p_{0}\overline{\eta}\left(  p\right)  \right] \label{eqn c9}%
\\
&  +i\theta^{2}g\delta_{ab}\int\frac{d^{4}p}{\left(  2\pi\right)  ^{4}}\left[
-2\left(  m^{2}+\overrightarrow{q}^{2}\right)  -2\left(  m^{2}+\overrightarrow
{p}^{2}\right)  +\left(  q_{0}\right)  ^{2}+\left(  p_{0}\right)  ^{2}\right]
\widetilde{\Delta}\left(  p\right) \nonumber
\end{align}

from eq$\left(  \ref{eqn c7}\right)  $ and eq$\left(  \ref{eqn c8}\right)  $
we get%

\begin{equation}
\widetilde{\Delta}_{ab}^{\left(  R\right)  }\left(  q\right)  =\widetilde
{\Delta}_{ab}\left(  q\right)  +\widetilde{\Delta}_{ac}\left(  q\right)
\Pi_{cd}^{\ast}\left(  q\right)  \widetilde{\Delta}_{db}\left(  q\right)
=\left[  \frac{1}{\widetilde{\Delta}^{-1}\left(  q\right)  -\Pi^{\ast}\left(
q^{2}\right)  -i\epsilon}\right]  _{ab}+O\left(  g^{2}\right)
\end{equation}

where $\Pi_{ab}^{\ast}\left(  q\right)  $ is the self-energy part (of the propagator)%

\begin{equation}
\Pi_{ab}^{\ast}\left(  q\right)  =\delta m^{2}\delta_{ab}-i\theta\delta
m^{2}q_{0}\varepsilon_{ab}-\frac{1}{4}\theta^{2}\delta m^{2}\left(
m^{2}+q^{2}\right)  \delta_{ab}-\frac{1}{4}\theta^{2}\delta m^{2}\left(
m^{2}+\overrightarrow{q}^{2}\right)  \delta_{ab}+\Pi_{\left(  Loop\right)
ab}^{\ast}\left(  q\right)
\end{equation}

Using the expression eq$\left(  \ref{eqn c9}\right)  $ of $\Pi_{\left(
Loop\right)  ab}^{\ast}\left(  q\right)  $, the self-energy part $\Pi
_{ab}^{\ast}\left(  q\right)  $ can be rewritten as%

\begin{align}
\Pi_{ab}^{\ast}\left(  q\right)   &  =\left[  \left(  1-\frac{1}{2}\theta
^{2}m^{2}\right)  \delta m^{2}\delta_{ab}+\pi_{ab}^{\left(  1\right)
}\right]  -\left[  i\theta\delta m^{2}\varepsilon_{ab}-\pi_{ab}^{\left(
2\right)  }\right]  q_{0}\label{eqn c10}\\
&  \ \ \ \ \ \ \ \ \ \ \ \ \ \ \ \ \ -\left[  \frac{1}{2}\theta^{2}\delta
m^{2}\delta_{ab}-\pi_{ab}^{\left(  3\right)  }\right]  \overrightarrow{q}%
^{2}+\left[  \frac{1}{4}\theta^{2}\delta m^{2}\delta_{ab}-\pi_{ab}^{\left(
4\right)  }\right]  \left(  q_{0}\right)  ^{2}\nonumber
\end{align}

where%
\begin{gather}
\pi_{ab}^{\left(  1\right)  }=4gi\int\frac{d^{4}p}{\left(  2\pi\right)  ^{4}%
}\left[  1-\frac{1}{2}\theta^{2}\left(  2m^{2}+\overrightarrow{p}^{2}\right)
+\frac{1}{4}\theta^{2}\left(  p_{0}\right)  ^{2}\right]  \widetilde{\Delta
}\left(  p\right)  \delta_{ab}-4g\theta^{2}\int\frac{d^{4}p}{\left(
2\pi\right)  ^{4}}p_{0}\overline{\eta}\left(  p\right)  \delta_{ab}%
\label{eqn c11}\\
\pi_{ab}^{\left(  2\right)  }=4g\theta\left[  \int\frac{d^{4}p}{\left(
2\pi\right)  ^{4}}\widetilde{\Delta}\left(  p\right)  \right]  \varepsilon
_{ab}\text{ \ , \ }\pi_{ab}^{\left(  3\right)  }=-2gi\theta^{2}\left[
\int\frac{d^{4}p}{\left(  2\pi\right)  ^{4}}\widetilde{\Delta}\left(
p\right)  \right]  \text{ \ , \ }\pi_{ab}^{\left(  4\right)  }=gi\theta
^{2}\left[  \int\frac{d^{4}p}{\left(  2\pi\right)  ^{4}}\widetilde{\Delta
}\left(  p\right)  \right]  \label{eqn c12}%
\end{gather}

the last three terms of eq$\left(  \ref{eqn c10}\right)  $ can be written as%

\begin{gather}
\left[  i\theta\delta m^{2}\varepsilon_{ab}-\pi_{ab}^{\left(  2\right)
}\right]  q_{0}=i\theta\left[  \delta m^{2}+4gi\int\frac{d^{4}p}{\left(
2\pi\right)  ^{4}}\widetilde{\Delta}\left(  p\right)  \right]  \varepsilon
_{ab}q_{0}\\
\left[  \frac{1}{2}\theta^{2}\delta m^{2}\delta_{ab}-\pi_{ab}^{\left(
3\right)  }\right]  \overrightarrow{q}^{2}=\frac{1}{2}\theta^{2}\left[  \delta
m^{2}+4gi\int\frac{d^{4}p}{\left(  2\pi\right)  ^{4}}\widetilde{\Delta}\left(
p\right)  \right]  \delta_{ab}\overrightarrow{q}^{2}\\
\left[  \frac{1}{4}\theta^{2}\delta m^{2}\delta_{ab}-\pi_{ab}^{\left(
4\right)  }\right]  \left(  q_{0}\right)  ^{2}=\frac{1}{4}\theta^{2}\left[
\delta m^{2}+4gi\int\frac{d^{4}p}{\left(  2\pi\right)  ^{4}}\widetilde{\Delta
}\left(  p\right)  \right]  \left(  q_{0}\right)  ^{2}%
\end{gather}

while the first term of eq$\left(  \ref{eqn c10}\right)  $ reads%

\begin{multline}
\left[  \left(  1-\frac{1}{2}\theta^{2}m^{2}\right)  \delta m^{2}\delta
_{ab}+\pi_{ab}^{\left(  1\right)  }\right]  =\left(  \delta m^{2}+4gi\int
\frac{d^{4}p}{\left(  2\pi\right)  ^{4}}\widetilde{\Delta}\left(  p\right)
\right)  \left[  \delta_{ab}-\frac{1}{2}\theta^{2}m^{2}\right] \\
+2gi\theta^{2}\left(  \int\frac{d^{4}p}{\left(  2\pi\right)  ^{4}}\left[
-m^{2}-\overrightarrow{p}^{2}+\frac{1}{2}\left(  p_{0}\right)  ^{2}\right]
\widetilde{\Delta}\left(  p\right)  +2gi\int\frac{d^{4}p}{\left(  2\pi\right)
^{4}}p_{0}\overline{\eta}\left(  p\right)  \right)  \delta_{ab}\nonumber
\end{multline}

Putting $\delta m^{2}=\delta m_{0}^{2}+\frac{1}{2}\theta^{2}\delta\mu_{0}^{2}$
in eq$\left(  \ref{eqn c10}\right)  ,$ we get%

\begin{equation}
\Pi_{ab}^{\ast}\left(  q\right)  =\left(  \delta m_{0}^{2}+4gi\int\frac
{d^{4}p}{\left(  2\pi\right)  ^{4}}\Delta\left(  p\right)  \right)  \pi
_{ab}^{\ast}\left(  q\right)  +\frac{1}{2}\theta^{2}\left[  \delta\mu_{0}%
^{2}\delta_{ab}+\varpi_{ab}^{\ast}\right]  \label{eqn c17}%
\end{equation}

where%

\begin{equation}
\Delta\left(  p\right)  =\frac{1}{m^{2}+p^{2}}%
\end{equation}

is the free-field propagator, while $\pi_{ab}^{\ast}\left(  q\right)  $\ and
$\varpi_{ab}^{\ast}$ are given by%

\begin{equation}
\pi_{ab}^{\ast}\left(  q\right)  =\left[  \delta_{ab}-i\theta\varepsilon
_{ab}q_{0}-\frac{1}{2}\theta^{2}\delta_{ab}\left(  m^{2}+\overrightarrow
{q}^{2}\right)  +\frac{1}{4}\theta^{2}\delta_{ab}\left(  q_{0}\right)
^{2}\right]  \label{eqn c18}%
\end{equation}

\begin{equation}
\varpi_{ab}^{\ast}=8gi\delta_{ab}\int\frac{d^{4}p}{\left(  2\pi\right)  ^{4}%
}\left[  -\frac{1}{4}+\frac{m^{2}+\overrightarrow{p}^{2}}{\left(  m^{2}%
+p^{2}\right)  ^{3}}\left(  p_{0}\right)  ^{4}\right]  \label{eqn c19}%
\end{equation}

\subsection{Dimensional Regularization}

To deal with the divergent integrals in eq$\left(  \ref{eqn c17}\right)  $ and
eq$\left(  \ref{eqn c19}\right)  $, one can use some sort of regularization
technique that makes these integrals finite. Dimensional regularization is the
most powerful and popular method of regulating divergent integrals in
perturbation theory. the idea is to treat the loop integrals as integrals over
D-dimensional momenta, then we could analytically continue the integrals back
to $D=4$. The ultraviolet divergences will then appear as singularities
(poles) in the deviation, $\epsilon=4-D$, from four dimensions \cite{QFT1}%
-\cite{QFT4} \cite{WEINB} \cite{BRN} \cite{Collins}.

Let us consider the divergent integral%
\begin{equation}
I_{1}=4gi\int\frac{d^{4}p}{\left(  2\pi\right)  ^{4}}\Delta\left(  p\right)
\end{equation}

The first step in doing this divergent integral is to analytically continue
the Feynman integrals to a continuous space-time dimension in the neighborhood
of the physical dimension $D=4$%

\begin{equation}
I_{1}=4gi\mu^{4-D}\int\frac{d^{D}p}{\left(  2\pi\right)  ^{D}}\frac{1}%
{m^{2}+p^{2}}%
\end{equation}

where $\mu$ is an arbitrary mass parameter introduced to keep the bare
coupling constant $g$ dimensionless.

The next step is called a Wick rotation \cite{WEINB} \cite{QFT2} \cite{QFT4},
instead of integrating $p^{0}$ on the real axis from $-\infty$ to $+\infty$,
we integrate it on the imaginary axis from $-i\infty$ to $+i\infty$. That is,
we can write $p^{0}=iq_{4}\ ,\ p^{2}=\overrightarrow{p}^{2}-\left(
p_{0}\right)  ^{2}=\overrightarrow{q}^{2}+\left(  q_{4}\right)  ^{2}=q^{2}$,
with $q_{4}$ integrated over real values from $-\infty$ to $+\infty$%

\begin{equation}
I_{1}=4gi\mu^{4-D}\int\frac{d^{D}p}{\left(  2\pi\right)  ^{D}}\frac{1}%
{m^{2}+p^{2}}=-4g\mu^{4-D}\int\frac{d^{D}q}{\left(  2\pi\right)  ^{D}}\frac
{1}{m^{2}+q^{2}}%
\end{equation}

Using the identity \cite{TOMS} \cite{Witt 2} \cite{QFT2} \cite{QFT4}%

\begin{equation}
a^{-z}=\frac{1}{\Gamma\left(  z\right)  }\int_{0}^{\infty}dtt^{z-1}e^{-at}
\label{eqn c22}%
\end{equation}

which is valid for $\operatorname{Re}z>0$ and $\operatorname{Re}a>0$, where
$\Gamma\left(  z\right)  $ is the Gamma function%
\begin{equation}
\Gamma\left(  z\right)  =\int_{0}^{\infty}dtt^{z-1}e^{-t} \label{eqn44}%
\end{equation}

defined for $\operatorname{Re}(z)>0$, we obtain the following expression%

\begin{equation}
I_{1}=-4g\mu^{4-D}\int\frac{d^{D}q}{\left(  2\pi\right)  ^{D}}\frac{1}%
{m^{2}+q^{2}}=-4g\frac{\mu^{4-D}}{\left(  2\pi\right)  ^{D}}\int_{0}^{\infty
}dt\int d^{D}qe^{-\left(  m^{2}+q^{2}\right)  t}%
\end{equation}

integration over the D-dimensional momentum integral can be performed by means
of the relation%
\begin{equation}
\int d^{D}qe^{-\left(  m^{2}+q^{2}\right)  t}=\left(  \frac{\pi}{t}\right)
^{\frac{D}{2}}e^{-m^{2}t}%
\end{equation}

Then we can perform the integral over $t$\ using the formula%
\begin{equation}
\int_{0}^{+\infty}dtt^{2s-1}e^{-\alpha t}=\frac{\alpha^{-s}}{2}\Gamma\left(
s\right)
\end{equation}

yielding to%

\begin{equation}
I_{1}=-4g\frac{\mu^{4-D}\pi^{\frac{D}{2}}}{\left(  2\pi\right)  ^{D}}\left(
m^{2}\right)  ^{\frac{D}{2}-1}\int_{0}^{\infty}dtt^{-\frac{D}{2}}%
e^{-t}=-4g\frac{m^{2}}{16\pi^{2}}\left(  \frac{m^{2}}{4\pi\mu^{2}}\right)
^{\frac{D}{2}-2}\Gamma\left(  1-\frac{D}{2}\right)
\end{equation}

The divergence of this integral manifests itself in the pole of the Gamma
function at the physical dimension $D=4$. In the neighborhood of this
dimension we set%

\begin{equation}
D=4-2s
\end{equation}

then%

\begin{equation}
I_{1}=-\frac{gm^{2}}{4\pi^{2}}\left(  \frac{m^{2}}{4\pi\mu^{2}}\right)
^{-s}\Gamma\left(  s-1\right)
\end{equation}

The singularity at small $s\rightarrow0$ may be isolated by using the expansions%

\begin{equation}
\Gamma\left(  s-1\right)  =-\left(  \frac{1}{s}-\gamma+1\right)  +O\left(
s\right)
\end{equation}

and%

\begin{equation}
\left(  \frac{m^{2}}{4\pi\mu^{2}}\right)  ^{-s}=1-s\ln\left(  \frac{m^{2}%
}{4\pi\mu^{2}}\right)  +O\left(  s\right)
\end{equation}

we get then%

\begin{equation}
I_{1}=4gi\mu^{4-D}\int\frac{d^{D}p}{\left(  2\pi\right)  ^{D}}\frac{1}%
{m^{2}+p^{2}}=\frac{gm^{2}}{4\pi^{2}}\frac{1}{s}+\frac{gm^{2}}{4\pi^{2}%
}\left[  1-\gamma-\ln\left(  \frac{m^{2}}{4\pi\mu^{2}}\right)  \right]
\end{equation}

Let us now evaluate the divergent integral eq$\left(  \ref{eqn c19}\right)  $%

\begin{equation}
I_{2}=8gi\int\frac{d^{4}p}{\left(  2\pi\right)  ^{4}}\left[  -\frac{1}%
{4}+\frac{m^{2}+\overrightarrow{p}^{2}}{\left(  m^{2}+p^{2}\right)  ^{3}%
}\left(  p_{0}\right)  ^{4}\right]
\end{equation}

dimensional regularization gives \
\begin{equation}
I_{2}=8gi\mu^{4-D}\int\frac{d^{D}p}{\left(  2\pi\right)  ^{D}}\left[
-\frac{1}{4}+\frac{m^{2}+\overrightarrow{p}^{2}}{\left(  m^{2}+p^{2}\right)
^{3}}\left(  p_{0}\right)  ^{4}\right]  =-8g\mu^{4-D}\int\frac{d^{D}q}{\left(
2\pi\right)  ^{D}}\left[  -\frac{1}{4}+\frac{m^{2}+\overrightarrow{q}^{2}%
}{\left(  m^{2}+q^{2}\right)  ^{3}}\left(  q_{4}\right)  ^{4}\right]  \
\end{equation}

where we have performed the usual Wick rotation.

Next we use the relation eq$\left(  \ref{eqn c22}\right)  $%

\begin{equation}
\frac{1}{\left(  m^{2}+q^{2}\right)  ^{3}}=\frac{1}{\Gamma\left(  3\right)
}\int_{0}^{\infty}dtt^{2}e^{-\left(  m^{2}+q^{2}\right)  t}%
\end{equation}

and the fact that \cite{QFT4}%

\begin{equation}
\int\frac{d^{D}q}{\left(  2\pi\right)  ^{D}}=0
\end{equation}

to get%

\begin{equation}
I_{2}=-8g\mu^{4-D}\int\frac{d^{D}q}{\left(  2\pi\right)  ^{D}}\frac
{m^{2}+\overrightarrow{q}^{2}}{\left(  m^{2}+q^{2}\right)  ^{3}}\left(
q_{4}\right)  ^{4}=-8g\frac{\mu^{4-D}}{\Gamma\left(  3\right)  }\int
\frac{d^{D}q}{\left(  2\pi\right)  ^{D}}\left(  m^{2}+\overrightarrow{q}%
^{2}\right)  \left(  q_{4}\right)  ^{4}\int_{0}^{\infty}dtt^{2}e^{-\left(
m^{2}+q^{2}\right)  t}%
\end{equation}

$I_{2}$ can be rewritten as%

\begin{equation}
I_{2}=-4g\frac{\mu^{4-D}}{\left(  2\pi\right)  ^{D}}\int_{0}^{\infty}%
dtt^{2}e^{-m^{2}t}\int d^{D-1}\overrightarrow{q}\left(  m^{2}+\overrightarrow
{q}^{2}\right)  e^{-\overrightarrow{q}^{2}t}\int_{-\infty}^{+\infty}%
dq_{4}\left(  q_{4}\right)  ^{4}e^{-q_{4}^{2}t}%
\end{equation}

the integration over the D-dimensional momentum integral on the right-hand
side can be performed with the help of the relations \cite{QFT4} \cite{TOMS}
\cite{Witt 2} \cite{QFT2}
\begin{equation}
\int d^{n}qf\left(  q^{2}\right)  =\frac{2\pi^{\frac{n}{2}}}{\Gamma\left(
\frac{n}{2}\right)  }\int_{0}^{+\infty}dkk^{n-1}f\left(  q^{2}\right)
\label{eqn c20}%
\end{equation}

and%

\begin{equation}
\int_{0}^{+\infty}dtt^{2s-1}e^{-\alpha t}=\frac{\alpha^{-s}}{2}\Gamma\left(
s\right)  \label{eqn c21}%
\end{equation}

hence%

\begin{equation}
I_{2}=-4g\frac{\mu^{4-D}}{\left(  2\pi\right)  ^{D}}\frac{2\pi^{\frac{D-1}{2}%
}}{\Gamma\left(  \frac{D-1}{2}\right)  }\Gamma\left(  \frac{5}{2}\right)
\int_{0}^{\infty}dtt^{-\frac{1}{2}}e^{-m^{2}t}\int_{0}^{+\infty}%
dkk^{D-2}\left(  m^{2}+k^{2}\right)  e^{-k^{2}t}%
\end{equation}

using eq$\left(  \ref{eqn c21}\right)  $, we get%

\begin{equation}
I_{2}=-g\frac{3\mu^{4-D}}{\left(  2\pi\right)  ^{D}}\pi^{\frac{D}{2}}\left(
m^{2}\right)  ^{\frac{D}{2}}\int_{0}^{\infty}dt\left[  t^{-\frac{D}{2}}%
+\frac{D-1}{2}t^{-\frac{D+2}{2}}\right]  e^{-t}%
\end{equation}

using again eq$\left(  \ref{eqn c21}\right)  $ to perform the integration over
$t$, leads to%

\begin{equation}
I_{2}=-g\frac{3m^{4}}{16\pi^{2}}\left(  \frac{m^{2}}{4\pi\mu^{2}}\right)
^{\frac{D}{2}}\left[  \Gamma\left(  -\frac{D}{2}+1\right)  +\frac{D-1}%
{2}\Gamma\left(  -\frac{D}{2}\right)  \right]
\end{equation}

The divergence of this integral manifests itself in the pole of the Gamma
function at the physical dimension $D=4$.

In the neighborhood of this dimension we set%

\[
D=4-2s
\]

then%

\begin{equation}
I_{2}=-g\frac{3m^{4}}{16\pi^{2}}\left(  \frac{m^{2}}{4\pi\mu^{2}}\right)
^{-s}\left[  \Gamma\left(  s-1\right)  +\left(  \frac{3}{2}-s\right)
\Gamma\left(  s-2\right)  \right]
\end{equation}

using the expansions%

\begin{align}
\Gamma\left(  s-1\right)   &  =-\left(  \frac{1}{s}-\gamma+1\right)  +O\left(
s\right) \\
\Gamma\left(  s-2\right)   &  =\frac{1}{2}\left(  \frac{1}{s}-\gamma+\frac
{3}{2}\right)  +O\left(  s\right)
\end{align}

and%
\begin{equation}
\left(  \frac{m^{2}}{4\pi\mu^{2}}\right)  ^{-s}=1-s\ln\left(  \frac{m^{2}%
}{4\pi\mu^{2}}\right)  +O\left(  s\right)
\end{equation}

we get the following expression%

\begin{equation}
I_{2}=\frac{3gm^{4}}{64\pi^{2}}\left[  \frac{1}{s}-\gamma+\frac{3}{2}%
-\ln\left(  \frac{m^{2}}{4\pi\mu^{2}}\right)  \right]
\end{equation}

Hence%

\begin{equation}
\varpi_{ab}^{\ast}=8gi\mu^{4-D}\delta_{ab}\int\frac{d^{D}p}{\left(
2\pi\right)  ^{D}}\left[  -\frac{1}{4}+\frac{m^{2}+\overrightarrow{p}^{2}%
}{\left(  m^{2}+p^{2}\right)  ^{3}}\left(  p_{0}\right)  ^{4}\right]
=\frac{3gm^{4}}{64\pi^{2}}\delta_{ab}\left[  \frac{1}{s}-\gamma+\frac{3}%
{2}-\ln\left(  \frac{m^{2}}{4\pi\mu^{2}}\right)  \right]
\end{equation}

and%

\begin{align}
\Pi_{ab}^{\ast}\left(  q\right)   &  =\left(  \delta m_{0}^{2}+\frac{gm^{2}%
}{4\pi^{2}}\frac{1}{s}+\frac{gm^{2}}{4\pi^{2}}\left[  1-\gamma-\ln\left(
\frac{m^{2}}{4\pi\mu^{2}}\right)  \right]  \right)  \pi_{ab}^{\ast}\left(
q\right) \\
&  \ \ \ \ \ \ \ \ \ \ \ +\frac{1}{2}\theta^{2}\left(  \delta\mu_{0}^{2}%
+\frac{3gm^{4}}{64\pi^{2}}\frac{1}{s}+\frac{3gm^{4}}{64\pi^{2}}\left[
\frac{3}{2}-\gamma-\ln\left(  \frac{m^{2}}{4\pi\mu^{2}}\right)  \right]
\right)  \delta_{ab}\nonumber
\end{align}

where%

\begin{equation}
\pi_{ab}^{\ast}\left(  q\right)  =\left[  \delta_{ab}-i\theta\varepsilon
_{ab}q_{0}-\frac{1}{2}\theta^{2}\delta_{ab}\left(  m^{2}+\overrightarrow
{q}^{2}\right)  +\frac{1}{4}\theta^{2}\delta_{ab}\left(  q_{0}\right)
^{2}\right]
\end{equation}

The condition that $m^{2}$ is the true mass of the particle is that the pole
of the propagator should (like the uncorrected propagator) be at $q^{2}%
=-m^{2}$, so that
\begin{equation}
\left[  \Pi_{ab}^{\ast}\left(  q\right)  \right]  _{q^{2}=-m^{2}}=\Pi
_{ab}^{\ast}\left(  \overrightarrow{0},m\right)  =0
\end{equation}

This condition allows us to evaluate $\delta m_{0}^{2}$ and $\delta\mu_{0}%
^{2}$%
\begin{align}
\delta m_{0}^{2}  &  =-\frac{gm^{2}}{4\pi^{2}}\frac{1}{s}-\frac{gm^{2}}%
{4\pi^{2}}\left[  1-\gamma-\ln\left(  \frac{m^{2}}{4\pi\mu^{2}}\right)
\right] \\
\delta\mu_{0}^{2}  &  =-\frac{3gm^{4}}{64\pi^{2}}\frac{1}{s}-\frac{3gm^{4}%
}{64\pi^{2}}\left[  \frac{3}{2}-\gamma-\ln\left(  \frac{m^{2}}{4\pi\mu^{2}%
}\right)  \right]
\end{align}

to this order $\Pi^{\ast}\left(  q\right)  $ is equal to zero%
\begin{equation}
\Pi^{\ast}\left(  q\right)  =0+O\left(  g^{2}\right)
\end{equation}

\section{Conclusion}

Thought this work we have considered a noncommutative complex scalar field
theory with self interaction, by imposing non commutativity to the canonical
commutation relations. The action and all relevant quantities are expanded up
to second order in the parameter of noncommutativity $\theta$. Using the path
integral formalism, the noncommutative free and exact propagators are
calculated to one-loop order and to the second order in the parameter of
noncommutativity $\theta$. Dimensional regularization was used to remove
ultraviolet divergences that arise from loop graphs. It has been shown that
these divergences may also be absorbed into a redefinition of the parameters
of the theory.

\end{document}